\documentclass[5p]{elsarticle}
\usepackage{lineno}
\usepackage{hyperref}
\usepackage{tabularx}
\usepackage{tabulary}
\usepackage{multirow}
\usepackage{amsmath}
\usepackage{algorithm}
\usepackage[noend]{algpseudocode}
\usepackage[toc,page]{appendix}
\usepackage{graphicx}
\usepackage{caption}
\usepackage{subcaption}
\usepackage{cleveref}
\makeatletter
\def\BState{\State\hskip-\ALG@thistlm}
\makeatother
\modulolinenumbers[25]
\journal{Big Data Research}
\begin{document}

\begin{frontmatter}

\title{Call Trace and Memory Access Pattern based \\ Runtime Insider Threat Detection for Big Data Platforms}

\author{Santosh Aditham, Nagarajan Ranganathan \& Srinivas Katkoori}
\address{4202 E. Fowler Avenue. Tampa, FL 33620-5399 U.S.A.}
\author{University of South Florida}
\ead[url]{http://csee.usf.edu}
\cortext[mycorrespondingauthor]{Corresponding author}
\ead{saditham@mail.usf.edu}

\begin{abstract}
Big data platforms such as Hadoop and Spark are being widely adopted both by academia and industry. In this paper, we propose a runtime intrusion detection technique that understands and works according to the properties of such distributed compute platforms. The proposed method is based on runtime analysis of system and library calls and memory access patterns of tasks running on the datanodes (slaves). First, the primary datanode of a big data system creates a \textit{behavior profile} for every task it executes. A behavior profile includes (a) trace of the system \& library calls made, and (b) sequence representing the sizes of private and shared memory accesses made during task execution. Then, the process behavior profile is shared with other replica datanodes that are scheduled to execute the same task on their copy of the same data. Next, these replica datanodes verify their local tasks with the help of the information embedded in the received behavior profiles. This is realized in two steps: (i) comparing the system \& library calls metadata, and (ii) statistical matching of the memory access patterns. Finally, datanodes share their observations for consensus and report an intrusion to the namenode (master) if they find any discrepancy. The proposed solution was tested on a small hadoop cluster using the default MapReduce examples and the results show that our approach can detect insider attacks that cannot be detected with the traditional analysis metrics.   
\end{abstract}


\end{frontmatter}

\linenumbers

\section{Introduction}
\begin{figure}
\centering
\includegraphics[scale=0.4]{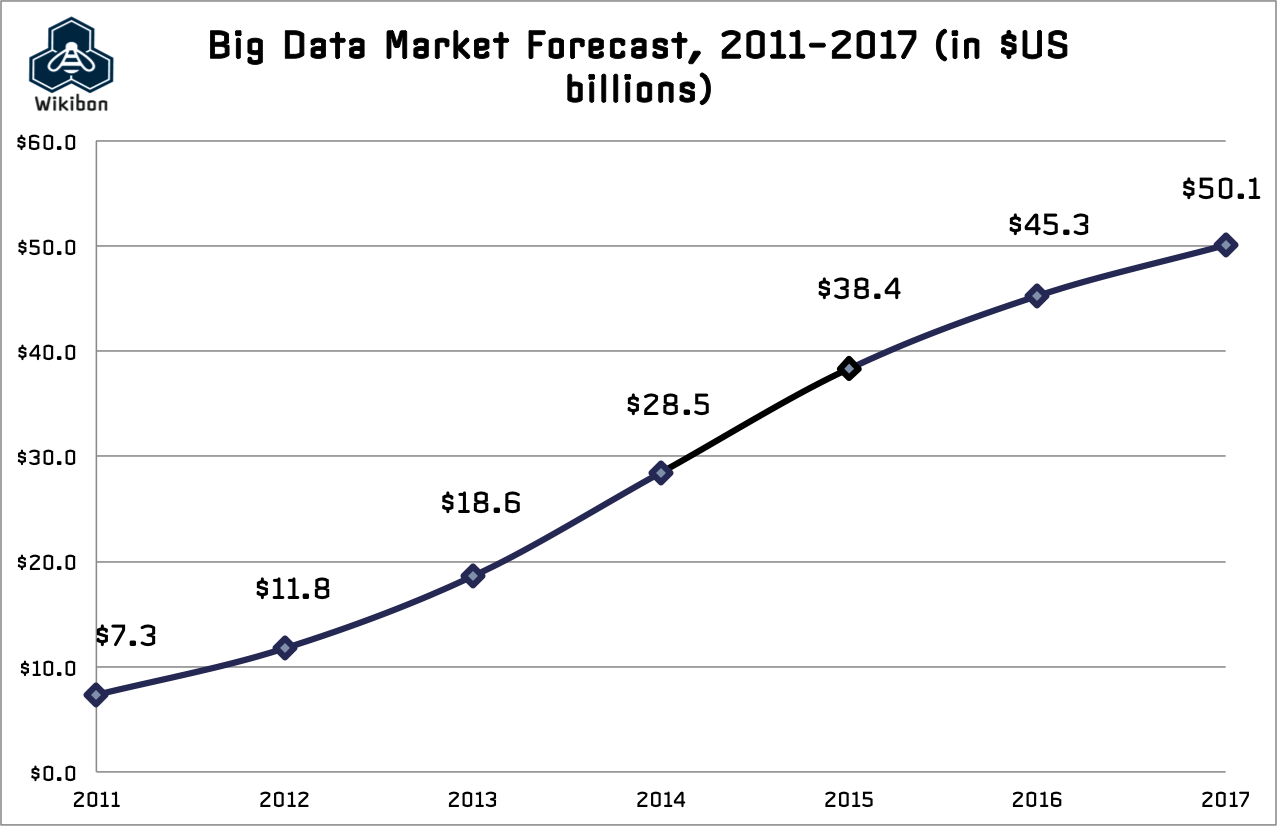}
\caption{Big Data Market Forecast \cite{forecast}}
\label{fig_forecast}
\end{figure}

The big data universe is growing aggressively with an estimated market of 50 billion dollars by next year. Figure \ref{fig_forecast} shows the increase in market forecast in the last five years. Big data platforms such as Hadoop \cite{hadoop} and Spark \cite{spark} are being widely adopted both by academia and industry. Security in these big data platforms is realized by incorporating some traditional security measures such as user authentication, access rights, activity logging and data encryption \cite{hadoop}. The end users have to trust the providers of big data platforms that host their data. Such trust is built on an underlying assumption that the platforms or their security methods will never be compromised. But unexpected issues such as insider attacks or control-flow attacks due to programmer errors can happen in any system anytime. 

Insider attacks (in an organization) typically deal with an employee stealing data using USB drives or by masquerading as another employee to gain access to unauthorized data \cite{santosh1}. Such attacks are hard to detect and almost impossible to prevent. But with the increase in popularity of concepts such as differential privacy in the big data universe, the biggest concern for these platforms is data loss or data being stolen and hence they need to be able to identify an attack on the data as soon as it happens. In this regards, we believe that big data platforms might not need brand new security methods but they need old methods to be applied in new combinations and with new emphasis. Our main focus is to identify such methods, modify them according to the platform needs and test them thoroughly before proposing to use them. In lieu of this, our previous works \cite{santosh1,santosh2,santosh3} address some of the data security concerns in big data platforms by proposing attack detection techniques that analyze compiled programs. But such static analysis techniques have their limitations. For example, they will not work if a datanode configuration is changed by an insider or if a rogue datanode intentionally masquerades the information it shares. Hence, in this work we propose a tighter intrusion detection technique that is based on runtime analysis of processes, especially their memory usage. This work is a full implementation of our position paper \cite{santosh4}.

%\subsection{Intrusion Detection at Compile-Time}
Typically in a big data cluster, when a user submits a request, the namenode (master) creates a job and schedules it for execution on the datanodes (slaves) that host the required data for the job. When the job is scheduled to execute at a datanode, static analysis techniques can be run on the associated compiled binary or bytecode to find vulnerabilities and bugs. Previously we proposed two such static analysis techniques that use control flow information \cite{santosh1,santosh2}. These static analysis methods for intrusion detection can help mitigate the effects of vulnerabilities caused due to misplaced jumps, uninitialized arguments, null pointers, dangling pointers (use after free), division by zero etc. 

%\subsection{Intrusion Detection at Run-Time}
Static analysis of a program binary helps in identifying some attacks but vulnerabilities due to buffer-overflows, shared libraries and dynamic linking will continue to exist even after static analysis. Memory corruption attacks can be detected and prevented at runtime with the help of sanitizers \cite{asan}. But handling improper calls due to programmer errors is still a difficult problem to address. Another difficult attack scenario to address is insiders and masqueraders in distributed environments such as cloud and big data \cite{insider}. Due to the distributed nature of big data platforms and their requirement to provide data consistency (with the help of replication), it is possible to perform dynamic analysis of processes for attack detection at runtime and still prevent adverse outcomes such as data loss. 

\subsection{Contributions of the proposed work}
In this paper, we propose a distributed, runtime intrusion detection technique that understands and works according to the properties of big data platforms. First, a behavior profile is created by the primary datanode for every job it executes. This profile contains system \& library call traces along with memory access patterns of the process as observed during process execution. The metadata from the call traces is used to verify consistency in call behavior among the datanodes. It can also be used for checking against normal behavior, if that information exists. As for memory access patterns, Principal Component Analysis (PCA) is used to observe orthogonal regression among multiple memory mapping aspects. The result of this analysis is a sequence that preserves the memory behavior of the datanode. Then, the process behavior profile consisting information about call and memory behavior of a datanode is shared with other replica datanodes that are scheduled to run the same job on the same data. Next, the replica datanodes verify the calls in their local process with the call information given in the received profiles. The replica datanodes also match the memory access pattern of their local process with that of the received, by using statistical analysis tests. Finally, datanodes share their observations for consensus before reporting an intrusion to the Namenode. 

The rest of this paper is organized as follows. Section II gives some background about big data platforms and runtime IDS. The various related works are also discussed here. Section III describes the security issues we address in this work. Section IV presents the proposed intrusion detection technique in detail. This section explains how call information and memory access patterns are gathered and matched for intrusion detection in a distributed system setup. Experimental setup and results are thoroughly discussed in Section V. Finally, Section VI draws the conclusions and outlines future work.

\section{Background and Related Work}
The four areas of security research that are closely related to this work are: (a) distributed intrusion detection techniques, (b) tracing application behavior for security in distributed systems, (c) handling vulnerabilities caused by programmer errors and protection from insider attacks, and (d) observing patterns in memory accesses made by a process for finding anomalies in process behavior. The other area of research used in this work is principal component analysis (PCA). 

\subsection{Distributed Intrusion Detection}
Intrusion detection systems (IDS) are used to detect anomalous or malicious usage in a computing device. Their design is based on one of two ways: (a) knowledge from prior attacks, and (b) learning from the behavior of programs and/or users. Knowledge-based IDS usually searches a program for known threat signatures that are stored in a database. With the drastic increase in the number of zero-day attacks, relying on a pre-populated database of threats is unsuitable. Even if we assume to have an ideal database of all possible threats, maintaining it would require a lot of resources and running search queries against it would be expensive. On the other hand, behavior-based IDS tries to model, analyze and compare user and/or application behavior to identify anomalies. Network usage (packets of data) is another common component observed by such IDS. This technique needs more resources and is more complex than signature-based IDS but it is more effective in a dynamically changing threat environment. Behavior-based IDS generally try to capture the context and apply statistics and rules on that context to detect anomalies. 

A distributed implementation of IDS is needed for systems that run on large clusters. Such an IDS would have centralized control and can detect behavioral patterns even in large networks. Efficient ways of data aggregation, communication and cooperation are key factors of success for such distributed IDS and it has to be employed at multiple levels: host, network and data \cite{cloudforensics}. Hence, using big data platforms to support general-purpose distributed IDS implementations is a recommended and popular practice. But in this work, we concentrate on building an IDS that can be used for security within a big data platform itself. IDS within a big data platform favors behavior-based distributed IDS because of the naturally large and ever increasing scope of threats.

\subsection{Tracing in Distributed Systems}
The need for tools that can diagnose complex, distributed systems is high because the root cause of a problem can be associated to multiple events/components of the system. Recent works in the distributed tracing domain are concentrating on providing an independent service. Magpie \cite{magpie} works by capturing events in the distributed system and uses a model-based system to store the traces. Xtrace \cite{xtrace} provides a comprehensive view for systems by reconstructing service behavior with the help of metadata propagation. Though it has similarities to our approach of providing task-centric causality, Xtrace concentrates on network level analysis. Retro \cite{retro} is another end-to-end tracing tool that audits resource usage along execution path. The drawback with tools such as Xtrace and Retro is that they are tightly coupled into the system and hence need the user to modify source code. HTrace \cite{htrace} is an Apache incubator project for distributed tracing which requires adding some instrumentation to your application. Pivot Trace \cite{pivottrace} is a dynamic causal monitoring service for distributed systems that provides a happened-before relation among discrete events. Fay \cite{fay} is another distributed event tracing platform that instruments code at runtime with safe extensions. $G^{2}$ \cite{g2} is a graph processing system for diagnosing distributed systems. Finally, \cite{provenance} proposed an anomaly detection method for MapReduce jobs in hadoop by collecting provenance data.

In this work, we are more interested in tracing system \&  library calls. Detecting intrusions using system call stack has been explored before. Hofmeyr et.al  \cite{syscall1} show that short sequences of system calls executed by running processes are a good discriminator between normal and abnormal operating characteristics. Many models such as Bayesian Classification, Hidden Markov Models (HMM) and process algebra have been proposed for system call sequence analysis \cite{syscall2,syscall3,syscall4,syscall5}. We use system and library call metadata to build behavior profile of a process. This is done by extracting information about system calls made during runtime from the call stack. Also, information related to library calls is included in our behavior profiles because big data frameworks use library calls that can be completely accounted for. This aspect of our approach is similar to AWS CloudTrail \cite{cloudtrail} which enables user to retrieve a history of API calls and other events for all of the regions within the user's account. Our work adopts the call trace anomaly based intrusion detection idea and modifies it to fit big data platforms accordingly. 

\subsection{Programmer Errors and Insider Attacks}
Programmer errors are a huge concern to security architects because anticipating the vulnerabilities due to programmer errors is difficult but at the same time they can give leeway to the attackers in their attempt to compromise a system. Generally, vulnerabilities due to programmer errors can be mitigated by enforcing control-flow integrity \cite{cfi}. More specifically, programmer errors that lead to memory corruptions can be handled by sanitizing memory instructions \cite{asan} in a program at compile-time. Though this approach is very memory expensive, it seems to work very efficiently for applications that run on a single machine. Practical usage on real-time distributed applications is not feasible. But in this work we concentrate on a subset of programmer errors that cannot be detected until runtime. One popular way of detecting attacks due to such vulnerabilities is by continuous monitoring of user and/or application behavior. This is a technique that is widely adopted by the industry in the recent past by leveraging big data platforms.

Insider attacks are known to be difficult to detect and prevent in general. This problem intensifies when the system under consideration is deployed on a large, distributed cluster. The ideal solution to detect and/or prevent insider attacks is by automating every aspect of a system such that there is no human intervention at all but obviously this is not feasible. Especially for big data systems, there is usually a service stack at the provider's end and another service stack at the client's end. Hence, cloud service providers such as Amazon and Google reduce the scope for insiders by adopting a two step procedure: (a) making most aspects of their systems to run automatically, and (b) asking their clients to do the same.

\subsection{Memory Access Patterns}
Understanding memory access patterns of big data applications will help in profiling them from their data usage perspective. Patterns in bandwidth usage, read/write ratio or temporal and spatial locality can be used when observing memory accesses of a process. For example, W.Wei et al. \cite{nvm} observed that memory access patterns of big data workloads are similar to traditional parallel workloads in many ways but tend to have weak temporal and spatial locality. One of the first works in characterizing memory behavior of big data workloads was done by Dimitrov et al. \cite{memory} who observed the characteristics such as memory footprints, CPI, bandwidth etc. of the big data workloads to understand the impact of optimization techniques such as pre-fetching and caching. In distributed compute systems, nodes of the cluster are typically virtual machines. For example, a Hadoop datanode is a process which dynamically dispatches tasks every time a job is scheduled. So, profiling the sizes of the private and shared memory accesses of all tasks will give the memory access pattern of the datanode.  

\subsection{Principal Component Analysis}
Principal Component Analysis (PCA) is an unsupervised linear transformation technique that finds the directions of maximal variance in a given dataset \cite{pca1,pca2,pcasvm}. A principal component is a linear combination of all the variables that retains maximal amount of information about the variables. When used for fitting a linear regression, PCA minimizes the perpendicular distances from the data to the fitted model. This is the linear case of what is known as orthogonal regression or total least squares \cite{pca1}, and is appropriate when there is no natural distinction between predictor and response variables. This is perfect for our memory access pattern matching problem because the features of memory access are all random and do not follow any predictor-response relation.

According to the theory of orthogonal regression fitting with PCA \cite{pca1}, $p$ observed variables can fit an $r$ dimensional hyperplane in $p$ dimensional space where $r$ is less than equal to $p$. The choice of $r$ is equivalent to choosing the number of components to retain in PCA. For this work, $r$ and $p$ are the same because we are not trying to reduce the dimensionality. But to profile memory usage of a process and later compare it with other profiles, we need a function that can explain the memory behavior. For this, we use the T-sqaured values that can be calculated using PCA in the full space.

\section{Threat Model}
A software-centric threat model is used in this work that dissects the big data platform design to understand its operational vulnerabilities. The primary focus of this work is to mitigate the effect of operational vulnerabilities caused at runtime due to improper usage of system and library calls by programmers. The reason for this choice is two-fold: (a) impact of operational vulnerabilities due to programmer errors cannot be estimated upfront, and (b) programmer errors are usually considered to be resolved at compile time. Our threat model also includes illegitimate use of data access privileges by insiders. For example, we want the proposed system to identify rogue datanodes masquerading as normal datanodes. For this purpose, we propose analyzing memory access patterns at process level. But it is difficult to differentiate between unusual accesses and corrupt accesses. For the scope of this work, we treat both scenarios as threats. Some assumptions about the system were made to fit this threat model. 
\begin{itemize}
\item All datanodes use the same architecture, operating system and page size. 
\item The path to framework installation is the same on all datanodes. 
\item The communication network will always be intact. 
\item The communication cost among replicas (slaves) is at most the communication cost between namenode (master) and datanodes (slaves). 
\end{itemize}

\section{Proposed Solution}
In this section, we discuss the proposed solution for detecting intrusions at runtime within the services provided by a big data platform. The motivation behind this work came from the requirement to have a runtime detection algorithm in the system architecture we had proposed previously \cite{santosh3}. Our framework is currently equipped with two compile time attack detection techniques \cite{santosh2,santosh3} proposed by us in the past. Hence, for this work it is safe to assume that the big data platform under consideration is equipped with our secure detection framework that can account for control-flow attacks that can be detected by static analysis. 

\subsection{Process Behavior Profile}
The first part of our proposed solution deals with building a behavior profile for every process running on a datanode. Process behavior profiles can be built by observing a variety of process characteristics. For this work, we are interested in designing the behavior of a process based on the system \& library calls and the memory accesses it makes during its runtime. While system and library calls help understand the work done by a process, memory accesses talk about the data usage of a process.

Our solution merges the information from these two aspects of a process - system \& library calls and memory access, to create a process behavior profile. This makes it hard for the attackers to masquerade. The algorithm for creating a process behavior profile is given in Algorithm \ref{alg_profile}. The process behavior profile representing the datanode will be a data structure with three entries: (1) identifier, (2) map with one entry per call, and (3) $t^2$, T-squared vector from PCA on memory access information. The identifier needs to be similar for a process across datanodes.

\subsubsection{System \& Library Calls}
A call instruction in a program has the potential to take the control away from program space to unknown territories. This property of a call instruction makes it an attractive target for attackers. In this work we focus on two specific kinds of call instructions known as the system calls and library calls. The system calls are programmatic ways of requesting a kernel service from the operating system. The list of possible system calls is specific to the operating system in use and the number of possible system calls is usually constant and limited. For example, Linux family of operating systems have approximately 140 system calls \cite{linux}. Since we specifically target big data platforms for this work, it is implicit that a certain framework such as hadoop or spark is installed on the cluster under surveillance. The advantage of library call monitoring is that the number of jars and shared library objects can be predetermined and these frameworks should have a predefined installation path which will not change unless there is a system level modification. 

The problem with tracing system \& library calls made by a process at runtime is that the order in which these calls are made might not persist across multiple runs of the same program. But this is important to our security framework since it tries to match information across replica datanodes. So simply trying to perform an exact match on the call stack will not work if we want to use call information for intrusion detection in a distributed computing domain. To combat this problem, we designed our process behavior profile to be descriptive of the calls made by a process. Instead of using the call stack, we extract metadata about system \& library calls from the call stack and use that information for intrusion detection. Each row in a process behavior profile representing a library or a system call describes it using four fields: (a) full class name of the callee, (b) method signature, and (c) source code line number and (d) count of the number of times this call was made by the process. A hash of the full class name is used as index for quick look-up. The other difficulty in using call information for intrusion detection is that the number of calls made by a process does not have to be the same for different datanodes. But a huge variation in the number of times a particular call is made can be used as an indicator for intrusion.

\begin{algorithm}[!t]
\caption{Algorithm to \textbf{create} process behavior profile}
\label{alg_profile}
\begin{algorithmic}[1]
\Procedure{Behavior Profile}{}
\State $\texttt{pid} \gets \text{get the process id of datanode}$
\State $\texttt{interval} \gets \text{set periodic interval for measurement}$
\BState \emph{getProfile(\texttt{pid})}:
\State $\texttt{Profile} \gets \text{empty map}$
\State $\texttt{Calls} \gets \textbf{call } {getCalls(\texttt{pid})}$
\State $\texttt{MemAccess} \gets \textbf{call } {getMemAccess(\texttt{pid})}$
\State $\texttt{Hash} \gets \text{hash of all call paths}$
\State $\texttt{Profile} \gets {insert(\texttt{[Hash, Calls], MemAccess)}}$
\State \textbf{return} $\texttt{Profile}$ 
\BState \emph{getCalls(\texttt{pid})}:
\While {$\texttt{callstack(pid)} = \text{system or library call}$}
\State $\texttt{callee} \gets \text{store the callee}$
\State $\texttt{signature} \gets \text{store the signature of the method}$
\State $\texttt{callPath} \gets \text{store the path}$
\State $\texttt{callCount} \gets \text{+1}$
\State $\texttt{hash} \gets \text{hash of the path}$
\State $\texttt{info} \gets \text{callee, signature, path, count}$
\EndWhile
\State \textbf{return} $\texttt{map(hash, info)}$
\BState \emph{getMemAccess(\texttt{pid})}:
\While{$\texttt{elapsed=interval}$}
\If {$\texttt{smaps(j).type} = \text{private or shared}$}
\State $\texttt{thisAccess[0]} \gets \texttt{smaps(j).Rss}$
\State $\texttt{thisAccess[1]} \gets \texttt{smaps(j).Private}$
\State $\texttt{thisAccess[2]} \gets \texttt{smaps(j).Shared}$
\EndIf
\State $\texttt{MemAccess} \gets add \texttt{ thisAccess}$
\EndWhile
\State $\texttt{Result} \gets \textbf{call }\texttt{PCA(MemAccess)}$
\State \textbf{return} $\texttt{Result}$
\EndProcedure
\end{algorithmic}
\end{algorithm}

\subsubsection{Memory Accesses}

While system \& library calls help in profiling a process and detect some attacks, they are susceptible to attacks as well. For example, a rogue datanode can masquerade its identity and send the process information to our security framework before an attack is launched. This will lead to a false negative scenario where the datanodes reach to a consensus about a process even though a rogue node compromised a process. Also, system calls in call stack give us the picture only until a file or device gets mapped to memory. All further read() and write() calls are made on the mapped memory using pointers. Hence, it is important to have an alternate perspective about the process in the behavior profile. Memory access information helps in the fine granularity of event reconstruction. Memory accesses during runtime give information about program characteristics such as the size of private and shared memory accessed by the program, number of clean and dirty pages in program memory etc. There are many advantages of using memory access patterns in behavior profiles, such as:
\begin{enumerate}[(1)]
\item information can be gathered periodically.
\item can be accomplished easily with hardware support.
\item gives insight about the data aspects of a process.
\item maintains differential privacy.
\end{enumerate}

\begin{figure*}[!t]
    \centering
    \begin{subfigure}[b]{0.3\textwidth}
        \includegraphics[trim = 15mm 60mm 15mm 60mm, clip, width=\textwidth]{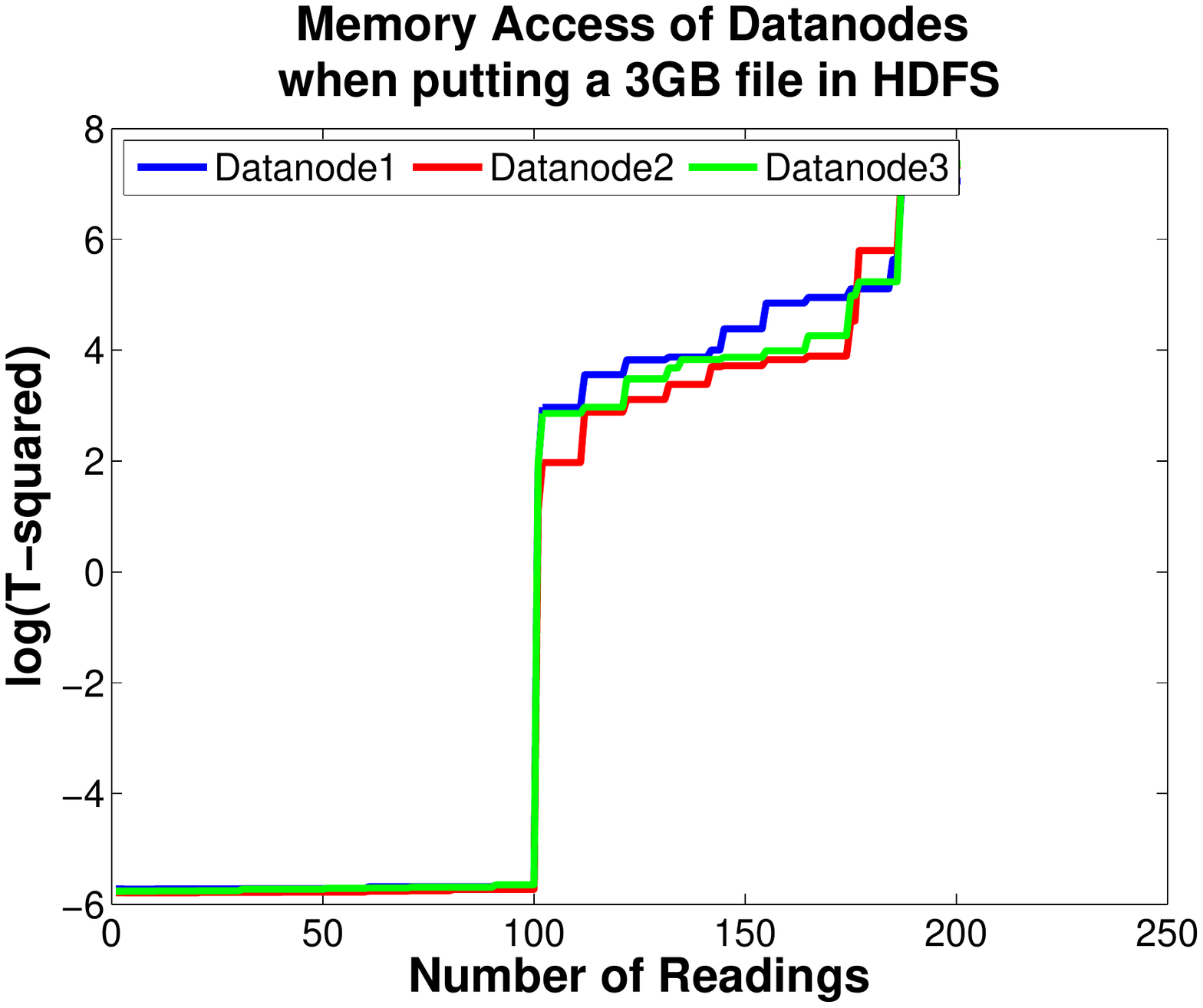}
        \caption{Memory Access Patterns of Datanodes when a 3GB file is \textbf{put} in HDFS}
        \label{fig:3gb}
    \end{subfigure}
    ~ %add desired spacing between images, e. g. ~, \quad, \qquad, \hfill etc. 
    \begin{subfigure}[b]{0.3\textwidth}
        \includegraphics[trim = 15mm 60mm 15mm 60mm, clip, width=\textwidth]{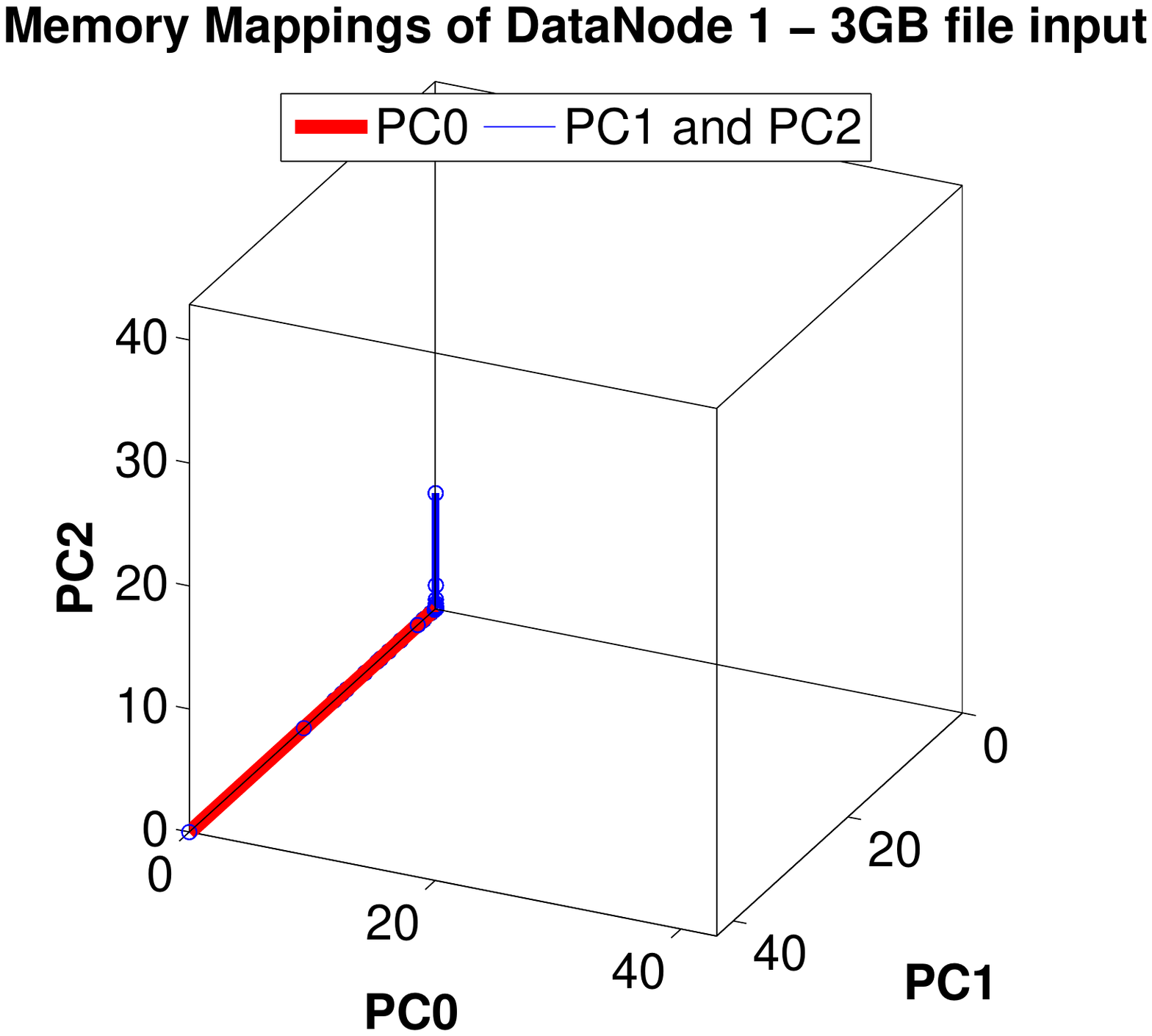}
        \caption{Orthogonal Regression among 3 features of Memory Access}
        \label{fig:regression}
    \end{subfigure}
    ~ %add desired spacing between images, e. g. ~, \quad, \qquad, \hfill etc. 
    \begin{subfigure}[b]{0.3\textwidth}
        \includegraphics[trim = 15mm 60mm 15mm 60mm, clip, width=\textwidth]{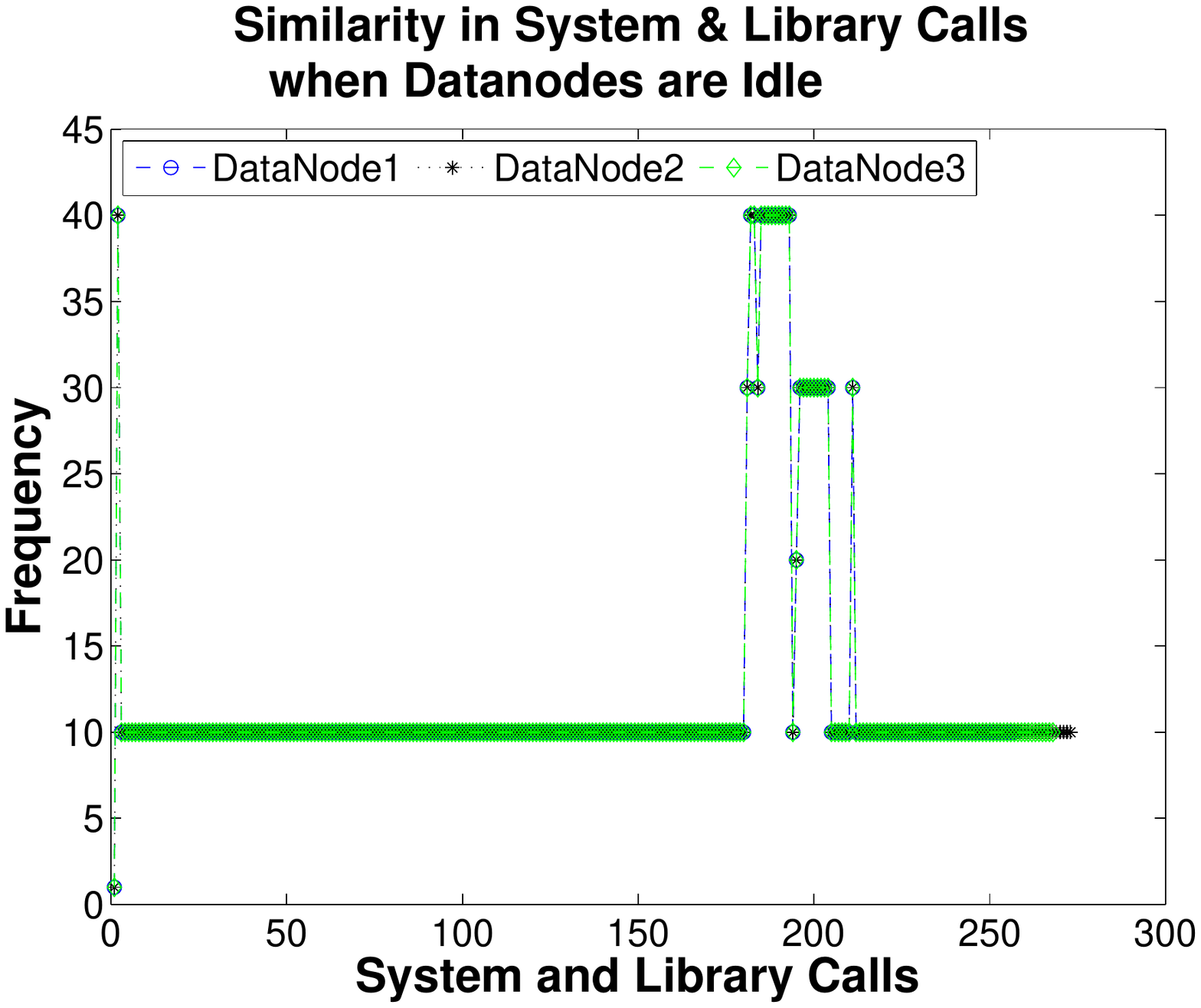}
        \caption{System \& Library Calls on Slave Nodes when cluster is \textbf{idle}}
        \label{fig:idle}
    \end{subfigure}
    \caption{Behavior of Datanodes in a Hadoop cluster} \label{fig:profile}
\end{figure*}

Today, most distributed systems are a cluster of nodes in their abstract forms i.e. each node is a virtual machine or a process running on a virtual machine. Hence, we designed our process behavior profile to include memory accesses made by the processes. The downside of this approach, especially with modern operating systems such as Linux, is that memory analysis becomes a complicated topic. It is extremely difficult to know about how memory is organized inside a running process and how the kernel handles the different allocation methods and process forking. For example, most modern operating systems use \textit{copy-on-write} semantics when forking a process where child process address space is mapped to the same backing pages (RAM) as the parent, except that when the child attempts to write to one of those pages, the kernel transparently copies the memory contents to a new, dedicated page, before carrying out the write. This approach speeds up the forking procedure but complicates the memory analysis. 

Usually, the kernel delays the actual allocation of physical memory until the time of the first access. Hence, knowing the actual size of physical memory used by a process (known as \textit{resident memory} of the process) is only known to the kernel. This memory mapping information from the kernel can be used to analyze the memory pattern of processes. A mapping is a range of contiguous pages having the same back-end (anonymous or file) and the same access modes. For this work, features of memory mapping that are relatively straight forward to analyze were chosen. The private and shared pages of a process in the RAM are observed as parts of memory access patterns. Private memory always belongs just to the process being observed while shared memory may be shared with its parent and/or children processes. In theory, the number of pages allocated to a process should be equal to the sum of its shared and private pages. To alleviate the penalty of constant monitoring, we gather this information periodically for every $2$ seconds. 

Two simple and typical big data work flows are used to demonstrate the insights provided by system calls and memory accesses of a process. The first example is about writing a 3GB file to HDFS in a Hadoop cluster. Figure \ref{fig:3gb} shows the results of principal component analysis on memory mappings ($t^2$) of the datanodes. The 3 dimension plot in Figure \ref{fig:regression} shows orthogonal regression among principal components which are calculated from observations made from memory mapping sizes of a datanode. The three dimensions used in this graph are the three different measurements taken at process level - resident set, private pages and shared pages. The red line indicates that the pages in RAM for a process are a combination of its private and shared pages. One observation or data point that seems to be an outlier. This can be due to multiple reasons such as swapping or giving away allocated memory to other processes in need etc. Table \ref{table_ftest_sample} has the results of f-test performed on $t^2$ statistic calculated as a result of PCA on the sample memory observations made during this test. A random sample of the smallest and largest memory accesses are taken into consideration for this test. Though this is atypical for statistical tests, the intent of this example is to show that the null hypothesis holds true. The first row in the table ($h = 0$) indicates the test accepts the null hypothesis. Here the null hypothesis is that the data comes from normal distributions with the same variance. The alternative hypothesis is that the population variance in memory access sizes of a datanode is greater than that of the other datanode. The second row of the table, $p$ values, are very high ($>0.5$) and imply confidence on the validity of the null hypothesis that variance in memory access is equal for all datanodes. The second example shows the simplest case with a hadoop cluster. Figure \ref{fig:idle} gives an insight to the system and library calls made by datanodes when they are idle i.e., no user submitted jobs but just maintenance. Each slave node made a total of 275 system calls during the time of observation. The calls and their frequencies were consistent across all data nodes. This can be observed with the overlapping call frequency patterns in Figure \ref{fig:idle}. Whether we consider the call information of idle nodes or the memory access information when putting a file in HDFS, it can be concluded that the datanodes are in harmony. 

\begin{table}
  \caption{Results of F-Test on Hotelling's T-squared statistic ($t^2$) from PCA when Datanodes are \textbf{idle}}
  \label{table_ftest_sample}
  \centering
  \begin{tabulary}{0.45\textwidth}{| C | C | C | C |} \hline
    \textbf{F-Test}&\textbf{Nodes \newline 1 \& 2}&\textbf{Nodes \newline 1 \& 3}&\textbf{Nodes \newline 2 \& 3} \\\hline
	h&	0		&	0		&	0 \\\hline
	p&	0.66	&	0.74	&	0.9 \\\hline
  \end{tabulary} 
\end{table}

Representing memory access pattern as a sequence of access size and using approximate string comparison or edit distance is one way to measure similarity between patterns. But there are many aspects to a memory access and creating a memory access profile with fine grained detail preserves more information for comparison across different machines. A straightforward comparison of all observed memory features is redundant and not always feasible. Hence, we chose to use an approximation method using multiple features when creating and comparing a process profile than using just one feature. In this work, each access pattern includes information about three features of a process memory access: (a) size of resident pages in the RAM for the mapping, (b) size of shared pages in that mapping, and (c) size of private pages in the same mapping. We use PCA to fit the measured 3 dimensional data as linear regression and share the resultant $t^2$  information for comparing and verifying memory access patterns of two datanodes. PCA calculates three metrics for a given sample dataset: \texttt{coefficients, scores and mean}. The coefficients of a principal component are given in descending order of component variance calculated by the singular value decomposition (SVD) algorithm \cite{pca2}. This is calculated using Equation \ref{eqCoeff} where $X$ is a principal component, $\lambda$ is the eigenvalue and $Y$ is the eigenvector. The sample means, $\bar{x}$ with $n$ observed memory access sizes per process is calculated using Equation \ref{eqMean} where $x_i$ is an individual memory access size from datanode $x$. The sample variances, ${\sigma^2}_x$ is calculated using Equation \ref{eqVariance} with $n-1$ degrees of freedom. Since our measurements use multiple features of a memory access, covariances are the eigenvalues of the covariance matrix of input data and they can be calculated using Equation \ref{eqCovariance} where $W_{x1, x2}$ is the covariance in two features of memory access of datanodes $x$. Here $x_{i,1}$ is the memory access size of the $i^{th}$ observation for the first memory feature. When observing $k$ memory features, we would have an array of values $[x_{i,1}, x_{i,2} ... x_{i,k}]$ for each observation. Scores are the representations of the input data in the principal component space. The $t^2$ values can be calculated from the memory patterns on each datanode using Equation \ref{eqHotelling}. But using PCA, the $t^2$ values are calculated as sum of squares distance from the center of the transformed space. Upon having the $t^2$ values, difference between them can be calculated using one way analysis of variance as given in Equation \ref{eqPCA} where the null hypothesis is that all group means are equal. Here, $t_x^2$ is the t-squared vector for datanode $x$, $t_y^2$ is the t-squared vector for datanode $y$ and $p$ is the probability that the means of $t_x^2$ and $t_y^2$ are the same.

\begin{equation}
\label{eqCoeff}
  X = \lambda ^ {1/2} Y
\end{equation}
\begin{equation}
\label{eqMean}
  \bar{x} = \frac{\sum_{i=1}^{n1}{x_i}}{n1}
\end{equation}
\begin{equation}
\label{eqVariance}
\begin{aligned}
  {\sigma^2}_x &= \frac {\sum_{i=1}^{n} (\bar{x} - x_i)}{n - 1} &
 \end{aligned}
\end{equation}
\begin{equation}
\label{eqCovariance}
\begin{aligned}
  W_{x1,x2} &= \frac{1}{n-1} \sum_{i=1}^{n} (x_{i,1} - \bar{x})(x_{i,2} - \bar{x})^{T} &
 \end{aligned}
\end{equation}
\begin{equation}
\label{eqHotelling}
 \begin{aligned}
  t_{x}^2 &= n (\bar{x} - \mu)^{T}W_x^{-1}(\bar{x} - \mu) &
 \end{aligned}
\end{equation}
\begin{equation}
\label{eqPCA}
 \begin{aligned}
  p &= anova(t_{x}^2 , t_{y}^2)
 \end{aligned}
\end{equation}

\begin{algorithm}[!t]
\caption{Algorithm to \textbf{verify} process behavior profile}
\label{alg_verify}
\begin{algorithmic}[1]
\Procedure{Verify Profile}{}
\State $\texttt{pid} \gets \text{get the process id from datanode}$
\State $\texttt{Local} \gets \text{behavior profile from this node}$
\State $\texttt{Recv} \gets \text{behavior profiles from other nodes}$
\BState \emph{compare()}:
\For{thread $\texttt{t}$ in $\texttt{pid}$}
\State $\texttt{result1} \gets \textbf{call }\texttt{CompareCalls(t)}$
\EndFor
\State $\texttt{result2} \gets \textbf{call }\texttt{CompareMemAccess(p)}$
\State $\texttt{result} \gets \texttt{result1 } \& \texttt{ result2}$
\State \textbf{notify} $\texttt{result}$ \Comment{similarity in calls \& memory accesses}
\BState \emph{CompareCalls(t)}:
\For{call $\texttt{c}$ in $\texttt{t}$}
\If {$\texttt{hash}(c_{path}) \mathrel{=} \texttt{Recv.find()}$}
\If {$\texttt{count}(c_{Local}) \mathrel{\ll} \mathrel{\gg} \texttt{count}(c_{Recv})$}
\EndIf
\State \textbf{return} $\texttt{true}$
\Else
\State \textbf{return} $\texttt{false}$
\EndIf
\EndFor
\BState \emph{CompareMemAccess(pid)}:
\If {$\texttt{compare(}\textit{$t_{Recv}^2$}, \textit{$t_{Local}^2$} \texttt{)}$} 
\State \textbf{return} $\texttt{true}$
\Else
\State \textbf{return} $\texttt{false}$
\EndIf
\EndProcedure
\end{algorithmic}
\end{algorithm}

\begin{algorithm}[!t]
\caption{Algorithm to \textbf{compare} process behavior profiles}
\label{alg_compare}
\begin{algorithmic}[1]
\Procedure{Compare Profiles}{}
\State $t^2_{Local} \gets \text{get the process profile from datanode}$
\State $t^2_{Recv} \gets \text{received process profiles}$
\ForAll {{$t^2_{i}$}}
\State $\texttt{filter}(t^2_{i})$ \Comment{remove tailing $t^2$ values}
\State $\texttt{sort}(t^2_{i})$
\EndFor
\If {$\texttt{Anova}(t^2_{Local}, t^2_{Recv})$}
\State $\texttt{compromised} \gets \texttt{Tukey}(t^2_{Local}, t^2_{Recv})$
\State \textbf{return} $\texttt{true}$
\Else
\State \textbf{return} $\texttt{false}$
\EndIf
\EndProcedure
\end{algorithmic}
\end{algorithm}

\begin{figure*}
\centering
\includegraphics[scale=0.55]{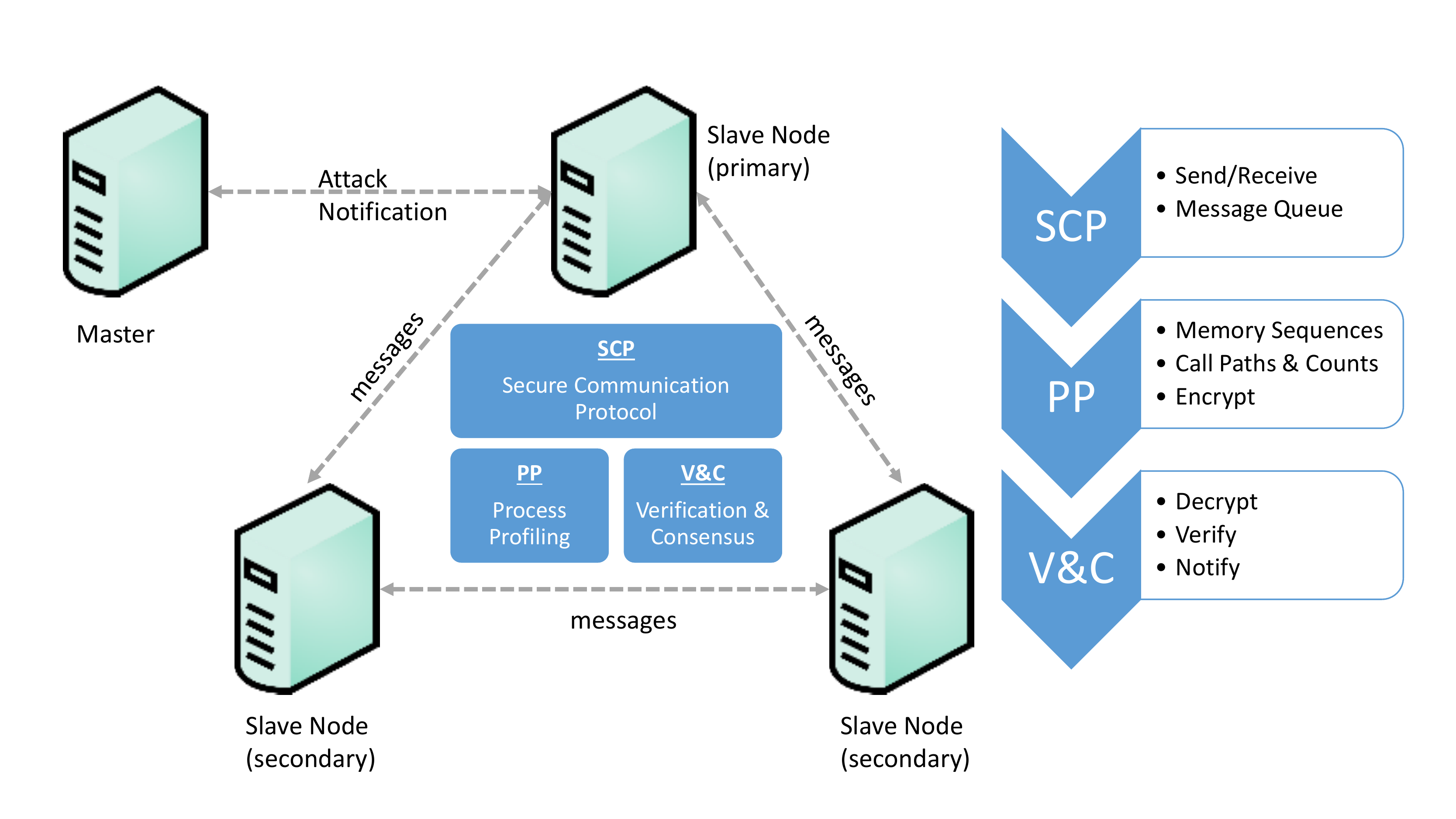}
\caption{A secure framework for hosting the proposed IDS \cite{santosh1}}
\label{fig_framework}
\end{figure*}

\subsection{Dynamic Verifier Function}
The dynamic verifier function is a part of the replica datanodes. It is used to parse a received behavior profile and use the extracted information to verify a local process. It will help in identifying process-level anomalies between two replica datanodes. We propose two algorithms as part of anomaly detection: (1) Algorithm \ref{alg_verify} is the generic verification algorithm that indicates an anomaly among process behavior profiles, and (2) Algorithm \ref{alg_compare} is the comparison algorithm for differentiating between two or more memory access patterns. The system \& library calls information is given in a \textit{hash map} data structure with call as the \textit{id} and call path as the \textit{value}. Finding differences at call path level is simple because the \textit{lookup()} function on the map will return the path in constant time. For every call made locally by a datanode, the call path is hashed using SHA-1 hashing algorithm and the hash map of calls received from the replica datanodes is looked up for the same hash in its index set. This lookup is quick and a mismatch (or) lack of match indicates that the datanodes used different set of system or library calls to perform the same task. This is a necessary but not a sufficient condition to indicate an intrusion. The additional information about calls available in the behavior profile helps in solidifying the attack detection process. The difference in the number of times a system or library call is called to perform the same task should be less than a predefined threshold, $\delta$, when comparing processes from different datanodes. 

Memory pattern of a process is represented using $t^2$ values of PCA. Since the $t^2$ values follow F-distribution, a comparison among memory patterns can be performed in two steps: (a) by running ANOVA test on the $t^2$ vectors to check if the patterns are different, and (b) by running Tukey test on the results from the ANOVA test to find the attacked datanode. This can also be accomplished by any other tests that assess the equality of variances such as F-test, Levene's test or Bartlett's test. In case of ANOVA, if the p-value is low ($< 0.05$) then it confirms the rejection of the null hypothesis with strong statistical significance. Then, a multiple comparison test such as a \textit{Tukey test} is used to check if the difference in the means of the $t^2$ values is significant. One big shortcoming of our approach is that it does not help in distinguishing between unusual process behavior from corrupt behavior. To be able to overcome such shortcomings, techniques such as reinforcement learning need to be used and we leave that for future work.

\subsection{System Architecture}
The proposed intrusion detection algorithm needs a strong framework to support it. For this purpose, we reuse the security framework we proposed previously for compile-time intrusion detection in big data platforms \cite{santosh1,santosh2}. Figure \ref{fig_framework} shows a high level design of the framework. This framework is equipped with an inter-node, key-based secure communication protocol. All the \textit{messages} among the datanodes are encrypted and use this communication protocol. The framework is assumed to be hosted on a coprocessor that communicates with the CPU for receiving the input data. Though a detailed design for such a coprocessor is not given, we believe that an ASIC based design would be a good idea for our coprocessor. It is assumed that the communication between the coprocessor and the main processor uses a secure protocol such as the one used by Apple processors to communicate with the secure enclave coprocessor \cite{conrad}. Adding new security instructions to the instruction set of a regular processor can also suffice. The other two elements of this framework are the process profiling phase and verification \& consensus phase. Algorithms \ref{alg_profile}, \ref{alg_verify} and \ref{alg_compare} are hosted and used for this purpose. The distributed nature of our algorithms help in conducting the profiling phase and the verification phase independently at each datanode in the cluster. This helps a lot in reducing the time taken for intrusion detection. Attack notification is sent from the primary datanode to the master node when there is a consensus among the datanodes about the existence of an attack. This consensus can be established using one of the popular leader election algorithms or consensus algorithms such as raft and paxos \cite{raft}. 

\begin{figure*}[!t]
    \centering
    \begin{subfigure}[b]{0.3\textwidth}
        \includegraphics[width=\textwidth, height=4cm]{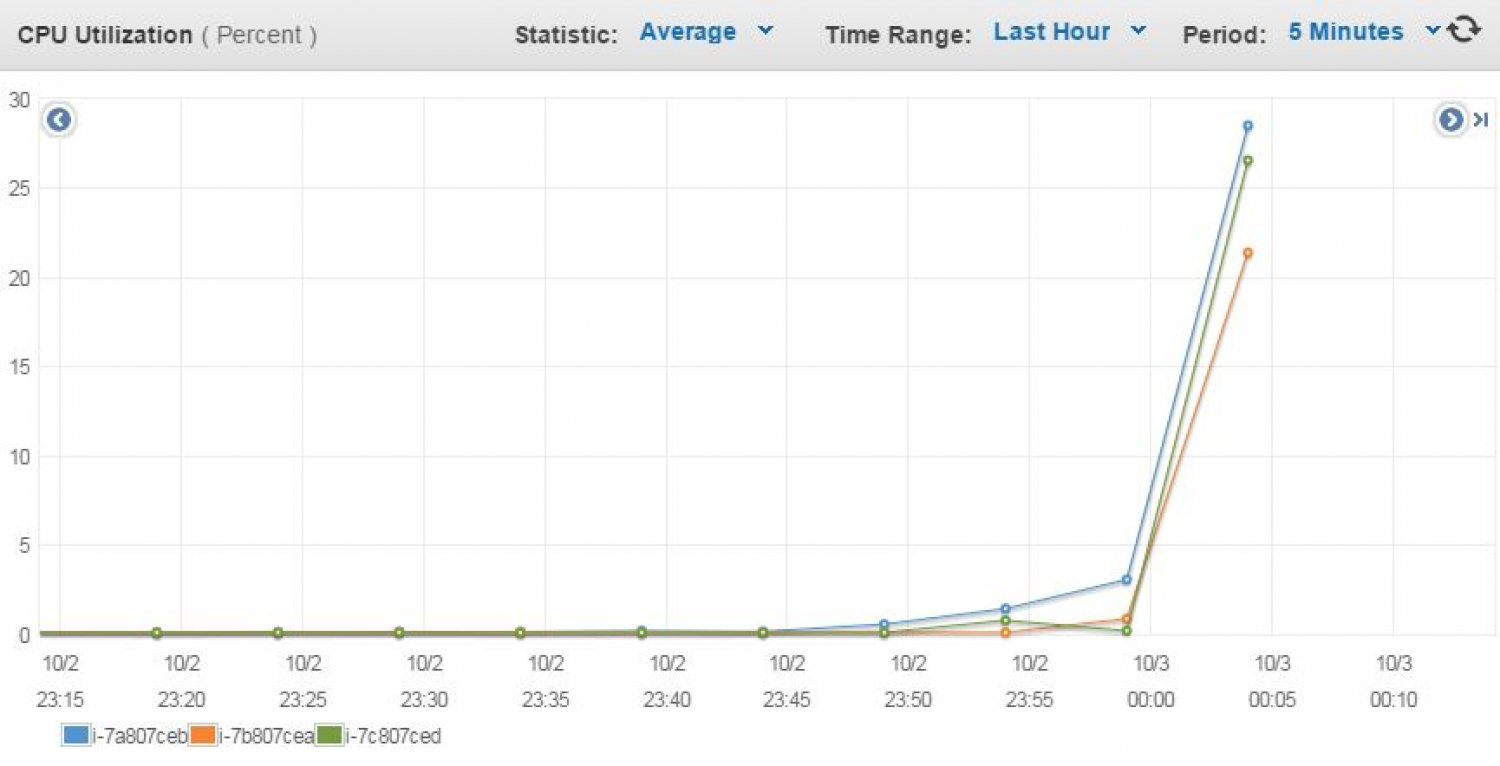}
        \caption{CPU utilization}
        \label{fig:cpu}
    \end{subfigure}
    ~ %add desired spacing between images, e. g. ~, \quad, \qquad, \hfill etc. 
    \begin{subfigure}[b]{0.3\textwidth}
        \includegraphics[width=\textwidth, height=4cm]{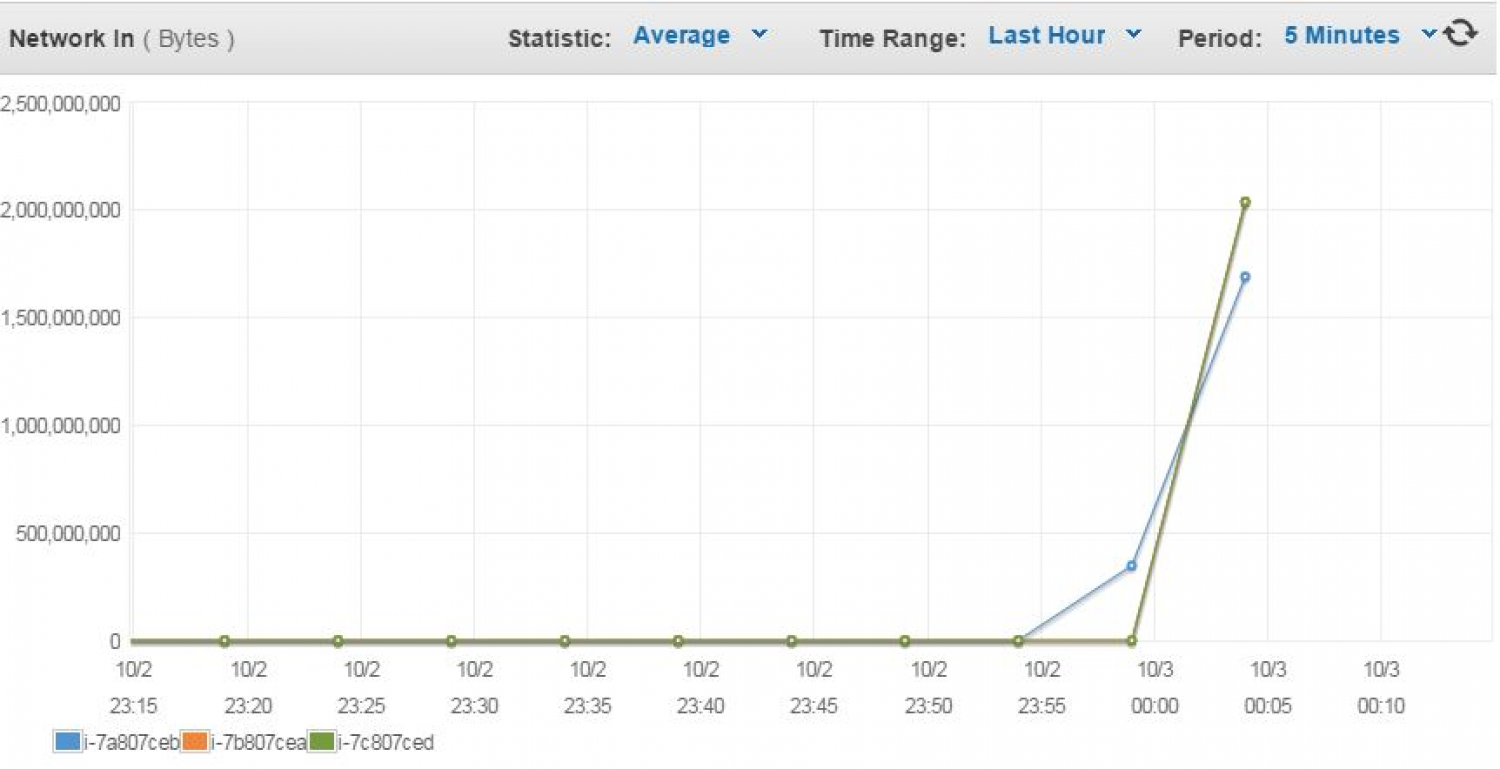}
        \caption{Network In}
        \label{fig:nwin}
    \end{subfigure}
    ~ %add desired spacing between images, e. g. ~, \quad, \qquad, \hfill etc. 
    \begin{subfigure}[b]{0.3\textwidth}
        \includegraphics[width=\textwidth, height=4cm]{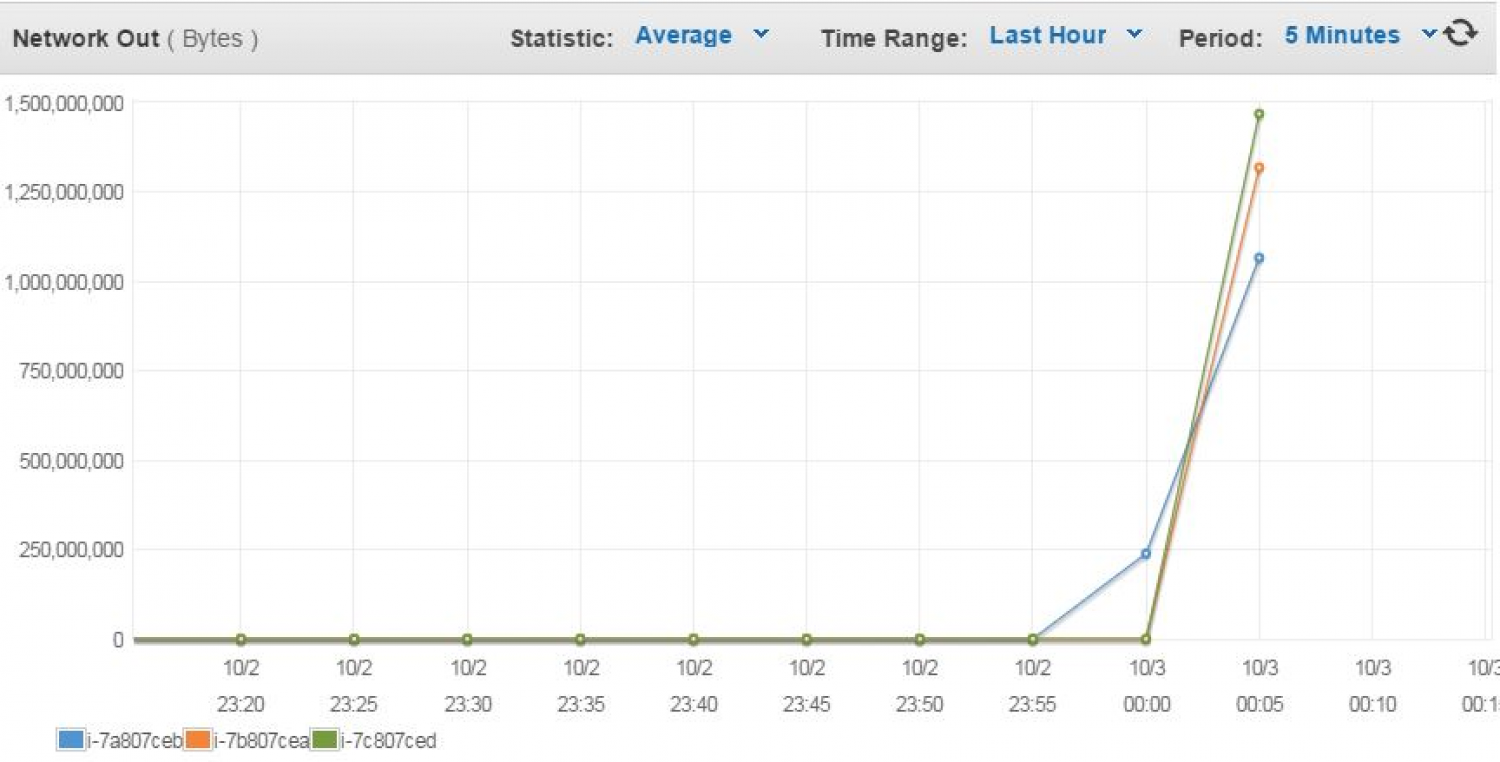}
        \caption{Network out}
        \label{fig:nwout}
    \end{subfigure}
    \caption{Results of monitoring the usual metrics of datanodes for \textbf{teragen} after a replica was compromised (source: Amazon EC2)}
    \label{fig:teragen}
\end{figure*}

\begin{figure*}[!t]
    \centering
    \begin{subfigure}[b]{0.3\textwidth}
        \includegraphics[width=\textwidth, height=4cm]{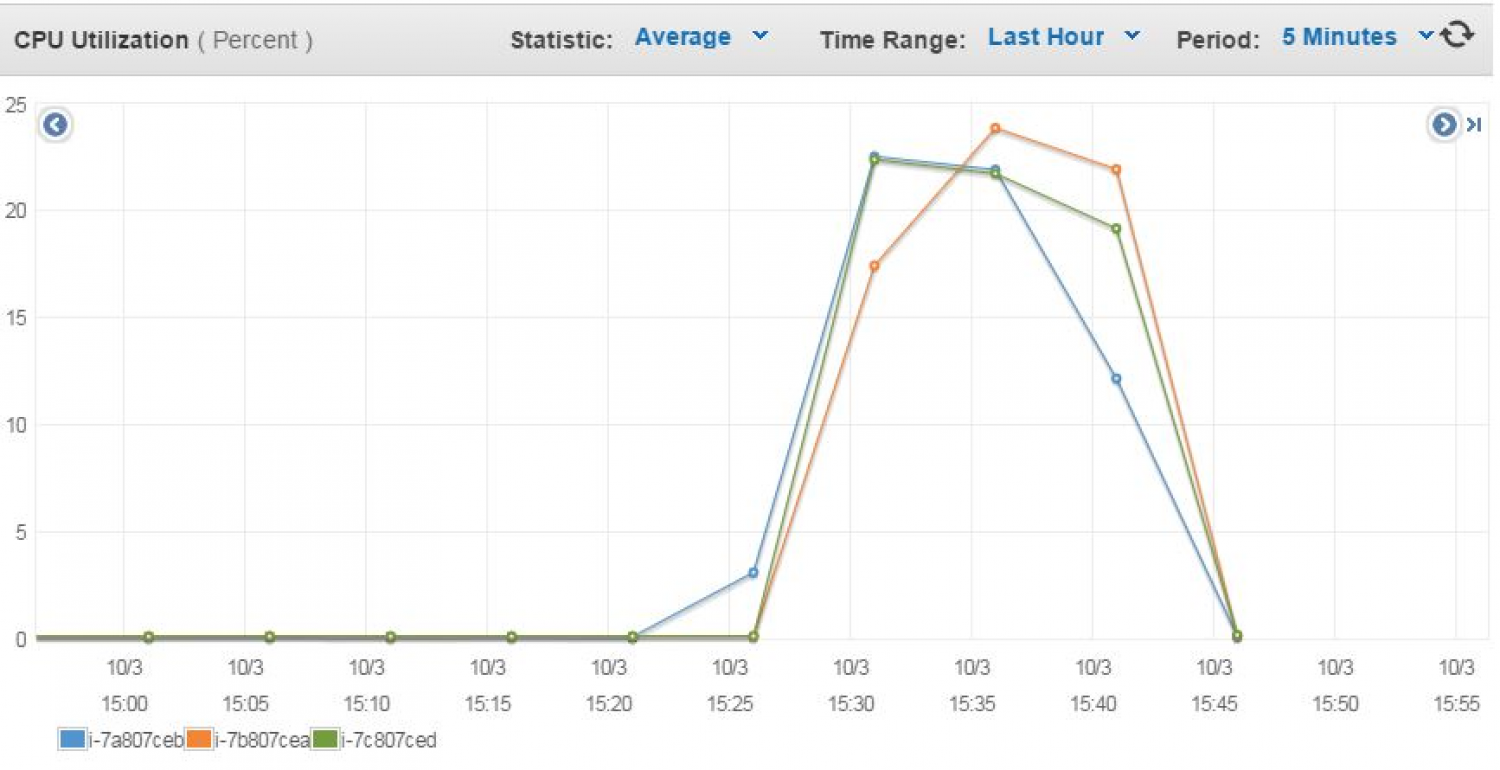}
        \caption{CPU utilization}
        \label{fig:cpu-ts}
    \end{subfigure}
    ~ %add desired spacing between images, e. g. ~, \quad, \qquad, \hfill etc. 
    \begin{subfigure}[b]{0.3\textwidth}
        \includegraphics[width=\textwidth, height=4cm]{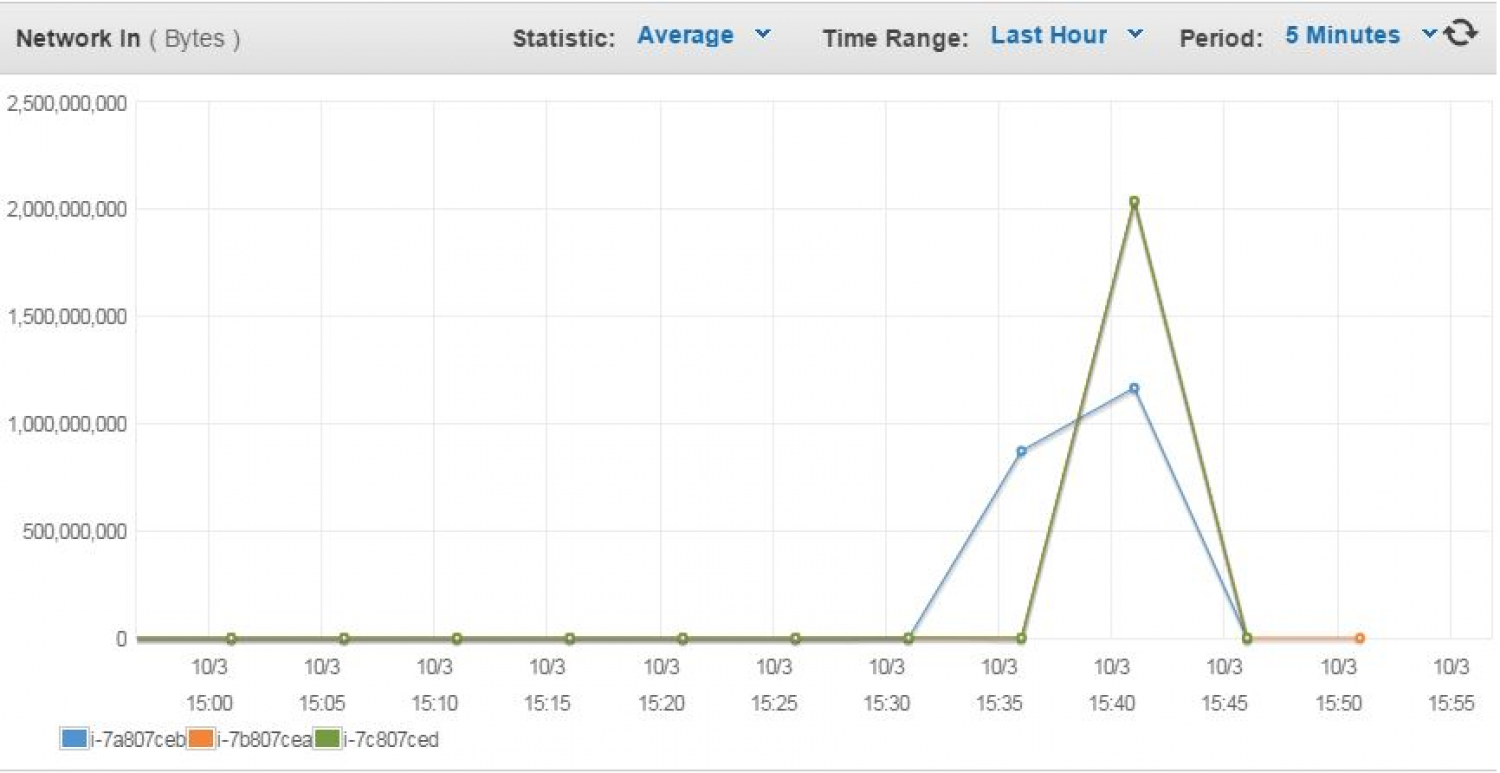}
        \caption{Network In}
        \label{fig:nwin-ts}
    \end{subfigure}
    ~ %add desired spacing between images, e. g. ~, \quad, \qquad, \hfill etc. 
    \begin{subfigure}[b]{0.3\textwidth}
        \includegraphics[width=\textwidth, height=4cm]{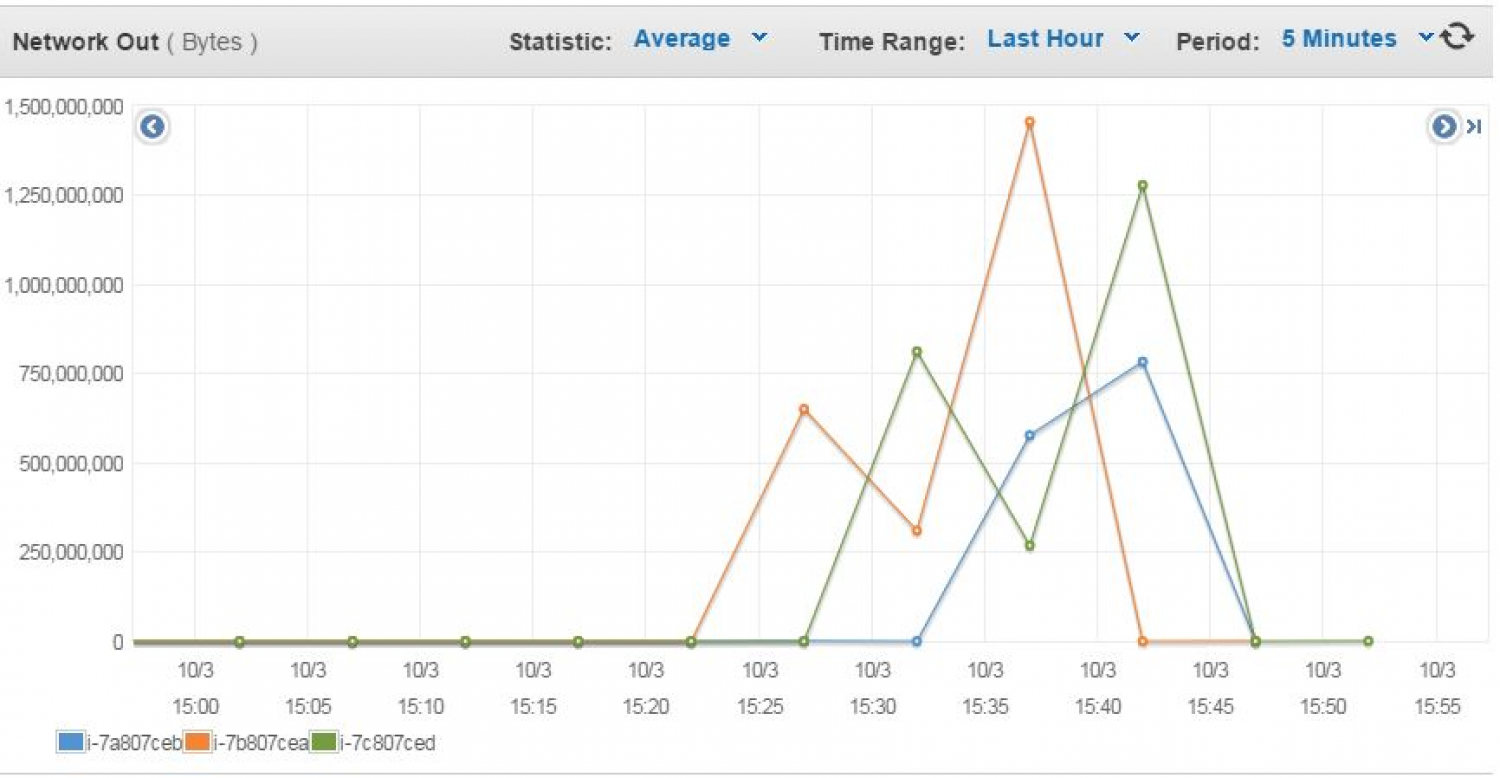}
        \caption{Network out}
        \label{fig:nwout-ts}
    \end{subfigure}
    \caption{Results of monitoring the usual metrics of datanodes for \textbf{terasort} after a replica was compromised (source: Amazon EC2)}
    \label{fig:terasort}
\end{figure*}

\section{Experiments and Results}
To test the proposed solution, we set-up a small Amazon EC2 cluster with 3 datanodes, 1 Namenode and 1 Secondary Namenode. Replication factor of the cluster is set to 3 (default). EC2 $m4.xlarge$ instances we used for hosting the cluster. Each node was running Ubuntu 14.04 and was equipped with a 2.4 GHz Intel Xeon® E5-2676 v3 (Haswell) processor, 4 virtual cores and 16 GB of memory.

In order to simulate a compromised cluster, one of the datanodes was explicitly programmed as the corrupt datanode. This was achieved by using two synthetic intrusions given in Table \ref{table_intrusions}. These synthetic intrusions represent different kinds of insider attacks such as: (a) misusing the system access privilege and modifying the system configuration, (b) misusing the data access privilege and copying user data for personal benefits, and (c) misusing the data access privilege and sharing or deleting sensitive user data as revenge against the system. Four of the sixteen hadoop examples that come by default with hadoop installation were used for demonstrating the results. A list of the MapReduce examples used along with a brief description is given in Table \ref{table_examples_hadoop}. Tests are conducted by running the hadoop map-reduce examples one at a time on the cluster. Observations from each data node are logged periodically (every 2 seconds) and later analyzed using our framework. Statistical analysis and graphs were generated using Matlab software \cite{matlab}. 

Two aspects of a process - system \& library calls and memory accesses are observed while running the Hadoop MapReduce examples on the cluster. The call stack of the process running on the data nodes is monitored. For library \& system call information, we get the path at the which the concerned jar file or shared library is located. For memory access pattern, we get the (a) memory footprint of a process by observing the number of pages referenced by the process, and (b) memory consumption of the process's mappings. The memory consumption can be calculated by looking at the the size of the mapping that is currently resident in RAM and the size of memory currently marked as referenced or accessed. In this work, we used the information available through \texttt{smaps} which only reports about memory pages that are actually in RAM. The memory consumptions of datanode processes are monitored by reading the values from \texttt{smaps} of all processes or tasks running on the datanode. There is a series of lines in the \texttt{smaps} file of a process for each mapping such as the following: \textit{Size, Rss, Pss, Shared Clean, Shared Dirty, Private Clean, Private Dirty, Referenced, Anonymous, KernelPageSize, MMUPageSize, Locked}. For proof of concept, we picked three of these features: \textit{Rss, Shared (clean and dirty), Private (clean and dirty)} because in theory Rss should sum up to the combined value of shares and private.

\begin{table}[!t]
  \caption{Two synthetic intrusions for testing our proposed solution}
  \label{table_intrusions}
  \centering
  \begin{tabulary}{0.45\textwidth}{|p{2.5cm}|p{4.5cm}|} \hline
    \textbf{Synthetic \newline Intrusion}&\textbf{Description} \\\hline
	Modify the \newline configuration & 				Change the configuration on one of the datanodes. For example, allocate less heap space to slowdown process execution. \\\hline
	Copy and share data & 				Access HDFS using a script and make unauthorized personal copies. Share the data using third party service like mail client.\\\hline
  \end{tabulary}
\end{table}

\begin{table}
  \caption{List of Hadoop Map-Reduce examples used in this work}
  \label{table_examples_hadoop}
  \centering
  \begin{tabulary}{0.45\textwidth}{| C | L |} \hline
    \textbf{Exp. Name}&\textbf{Description} \\\hline
	Random text writer & 	A map/reduce program that writes 10GB of random textual data per node. \\\hline
	Aggregate word count & 	An Aggregate based map/reduce program that counts the words in the input files. \\\hline
	Teragen & 				Generate one terabyte of randomly distributed data. \\\hline
   	Terasort & 				Sort one terabyte of randomly distributed data. \\\hline
  \end{tabulary}
\end{table}

\subsection{Experiments}
A common problem for big data related academic researchers is the relative lack of high-quality intrusion detection data sets \cite{nodataset}. This is a much bigger problem if the attacks under consideration are not network related. So, we decided to create our own synthetic attacks. Once the system was setup, two synthetic insider attacks were performed on system while it was executing the four Hadoop MapReduce examples to emulate normal usage of the services. 
\subsubsection{Attack 1: Modifying the node configuration}
This attack involves exploitation of access privileges by an insider who is legally allowed to access the system and its configuration files. An insider who is a system admin can modify the configuration properties of a datanode to intentionally impact the performance of the overall system. To implement this attack, we made the system admin change the datanode configuration through the \texttt{hdfs-site.xml} file on of the datanodes of the hadoop cluster. The amount of memory allocated for non-DFS purposes on the datanode were increased by 25\% and the number of server threads for the datanode were reduced by changing the handler count to 2. Since this is a one-time modification made by an authorized user whose job entails modification of the configuration files, usual user-profiling will not help in detecting the attack.  

\subsubsection{Attack 2: Illicit copying of data}
This attack involves two cases: (a) the use of non-certified (and untrusted) equipment to transfer data from one machine to another, and (b) the use of certified and trusted software (such as a mail client) to transfer data from one machine to another. Similar to the previous attack, the first step involved in this attack is for the system admin to modify the configuration through the \texttt{hdfs-site.xml} file on of the datanodes of the Hadoop cluster. A new location local to the system admin account is added to the DFS data directory property. As a result, all blocks at this datanode have two copies - one copy in the actual HDFS location used while setting up the cluster and another duplicate copy in the system admin's local folder. Next, a script is used to simulate an insider periodically transferring these duplicates files from his local folder of  to another remote location using the mail client service or USB device. Since it is not possible for us to connect a USB device to Amazon EC2 instances, we included the system calls involved with using such a device in the attack script.

\subsubsection{Results from Hadoop}
The results of the Hadoop MapReduce examples are given in Table \ref{table_basic_results}. Terasort and Teragen examples were run on a terabyte of data while Random text writer and aggregate word counter used a little more than 10GB of data. Because of this variation in data size, it can be noticed that the time taken to complete these examples also changed accordingly. To generate the terabyte of input data, Teragen took 109 seconds while Terasort took more than 6 times that amount (695 seconds) to sort the terabyte of data. Random text writer took 22.5 seconds to generate random words of size 10GB and Aggregate word count took just 14 seconds to count the words in that 10GB of input data.

\subsubsection{Results from Proposed Method}
While the hadoop MapReduce examples were executing the way they are supposed to, our security framework performed its analysis on the datanodes that were contributing to the successful execution of those MapReduce examples.

\begin{table*}[!t]
  \caption{Information and experimental results for Hadoop MapReduce examples}
  \label{table_basic_results}
  \centering
  \begin{tabulary}{1\textwidth}{|L|C|C|CCC|C|C|C|} \hline
    \multirow{ 2}{*}{\textbf{Example}}& \multirow{ 2}{*}{\textbf{Data Size}}& \multirow{ 2}{*}{\textbf{Time}}& \multicolumn{3}{c|}{\textbf{No. of observations}}& \multirow{ 2}{*}{\textbf{Sum of}}& \multirow{ 2}{*}{\textbf{F}}& \multirow{ 2}{*}{\textbf{p-value}} \\
    &	(Bytes)&	(seconds)&	\textit{Node1}&	\textit{Node2}&	\textit{Node3}&	\textbf{Squares}&	\textbf{Statistic}& \\\hline
	Teragen&	10000000000&	109.313&	58770&	59970&	60114&	0.129&	102.95& 2.1 $e^{-45}$\\\hline
	Terasort&	10000000000&	694.966&	118940&	127310&	124088&	0.256&	162.19& 3.9 $e^{-71}$\\\hline
	Random Text Writer&		1102236330&		22.543&		29681&	31850&	31025&	0.094&	48.64& 7.7 $e^{-22}$\\\hline
	Aggregate Word Count&		1102250820&		14.347&		29675&	31850&	31157&	0.069&	37.29& 6.4 $e^{-17}$\\\hline
  \end{tabulary}
\end{table*}

\begin{figure*}[!t]
    \centering
    \begin{subfigure}[b]{0.23\textwidth}
        \includegraphics[trim = 15mm 70mm 15mm 60mm, clip, width=\textwidth, height=5cm]{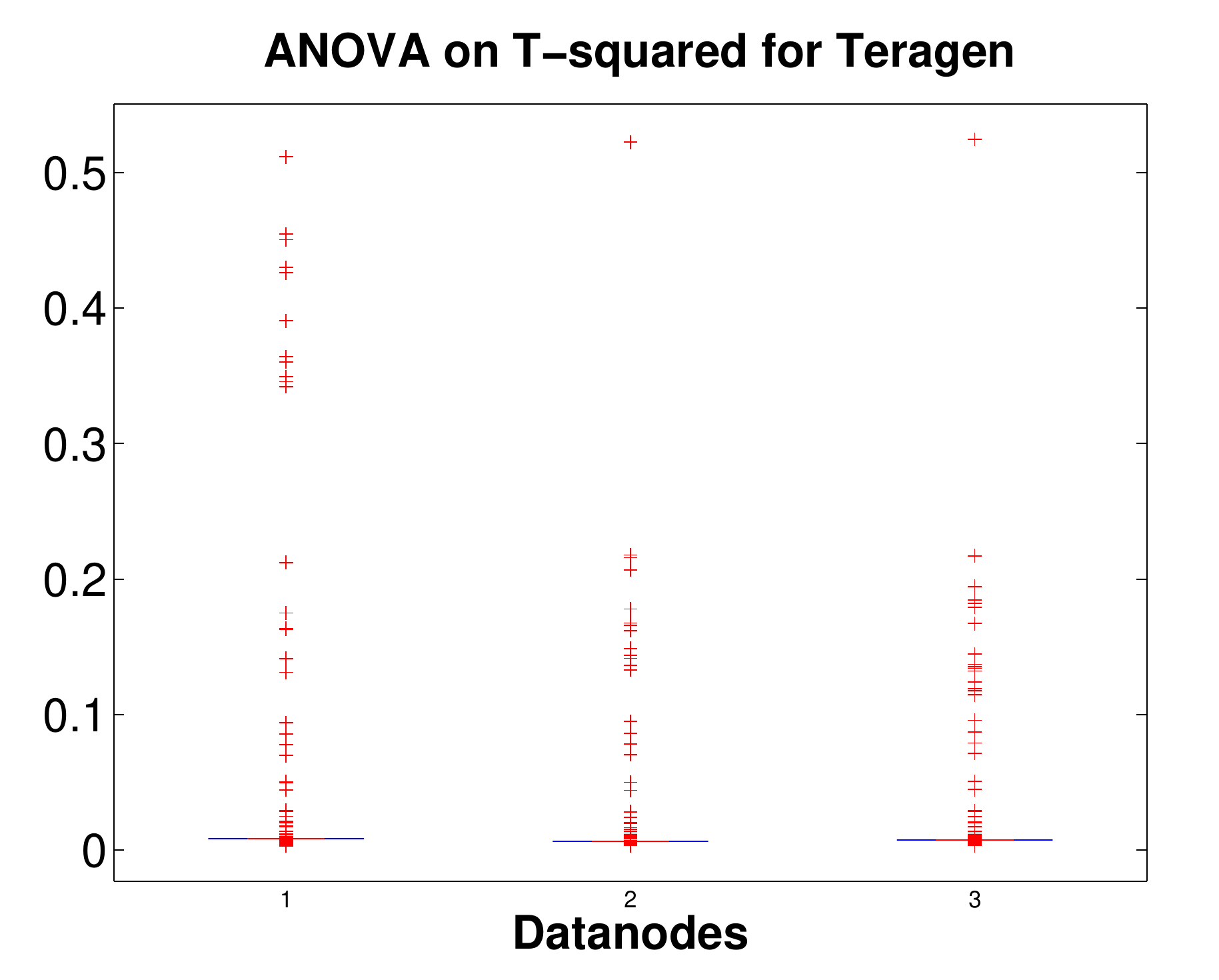}
        \caption{Teragen}
        \label{fig:anova-tg}
    \end{subfigure}
    ~ 
    \begin{subfigure}[b]{0.23\textwidth}
        \includegraphics[trim = 15mm 70mm 15mm 60mm, clip, width=\textwidth, height=5cm]{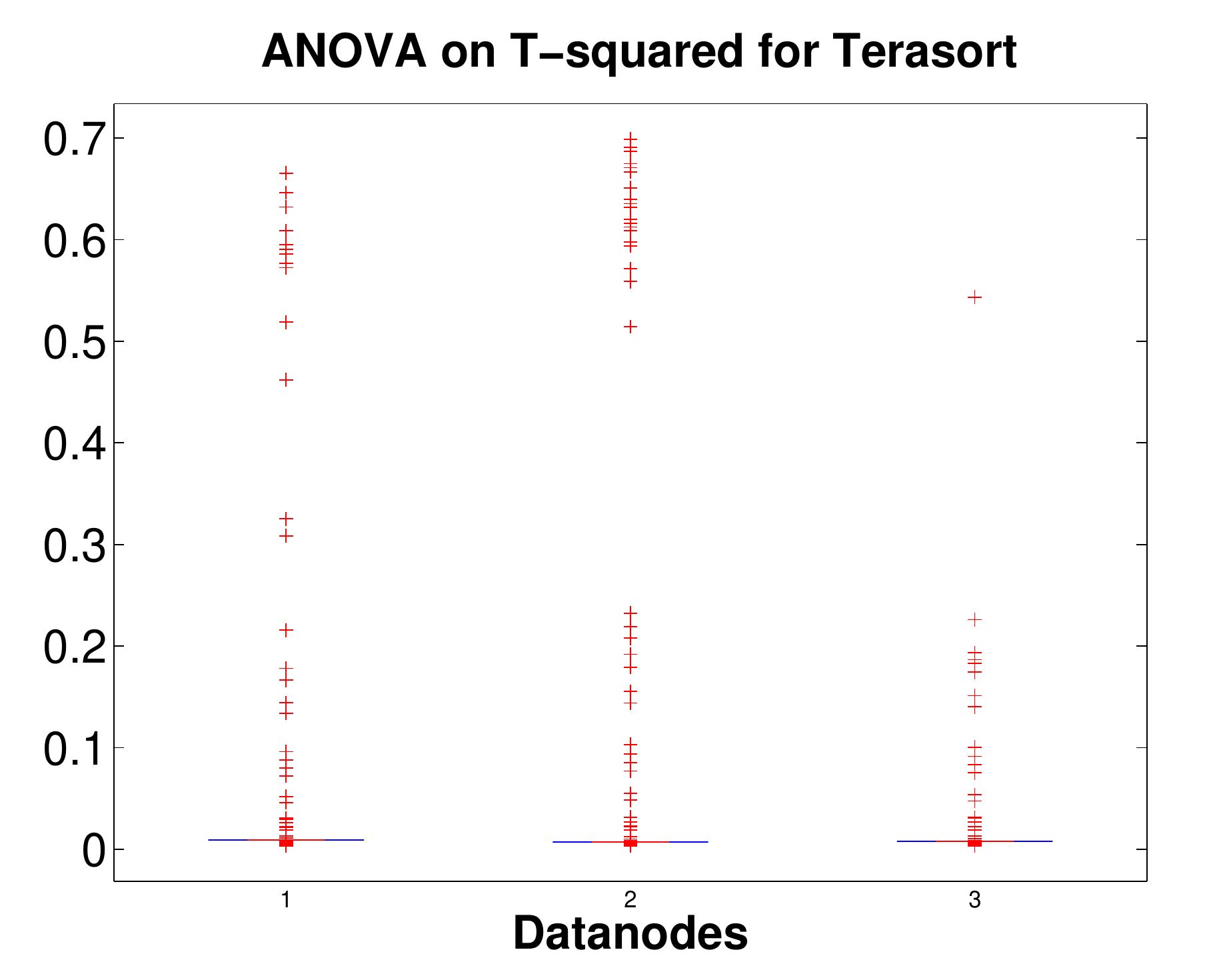}
        \caption{Terasort}
        \label{fig:anova-ts}
    \end{subfigure}
    ~ 
    \begin{subfigure}[b]{0.23\textwidth}
        \includegraphics[trim = 15mm 70mm 15mm 60mm, clip, width=\textwidth, height=5cm]{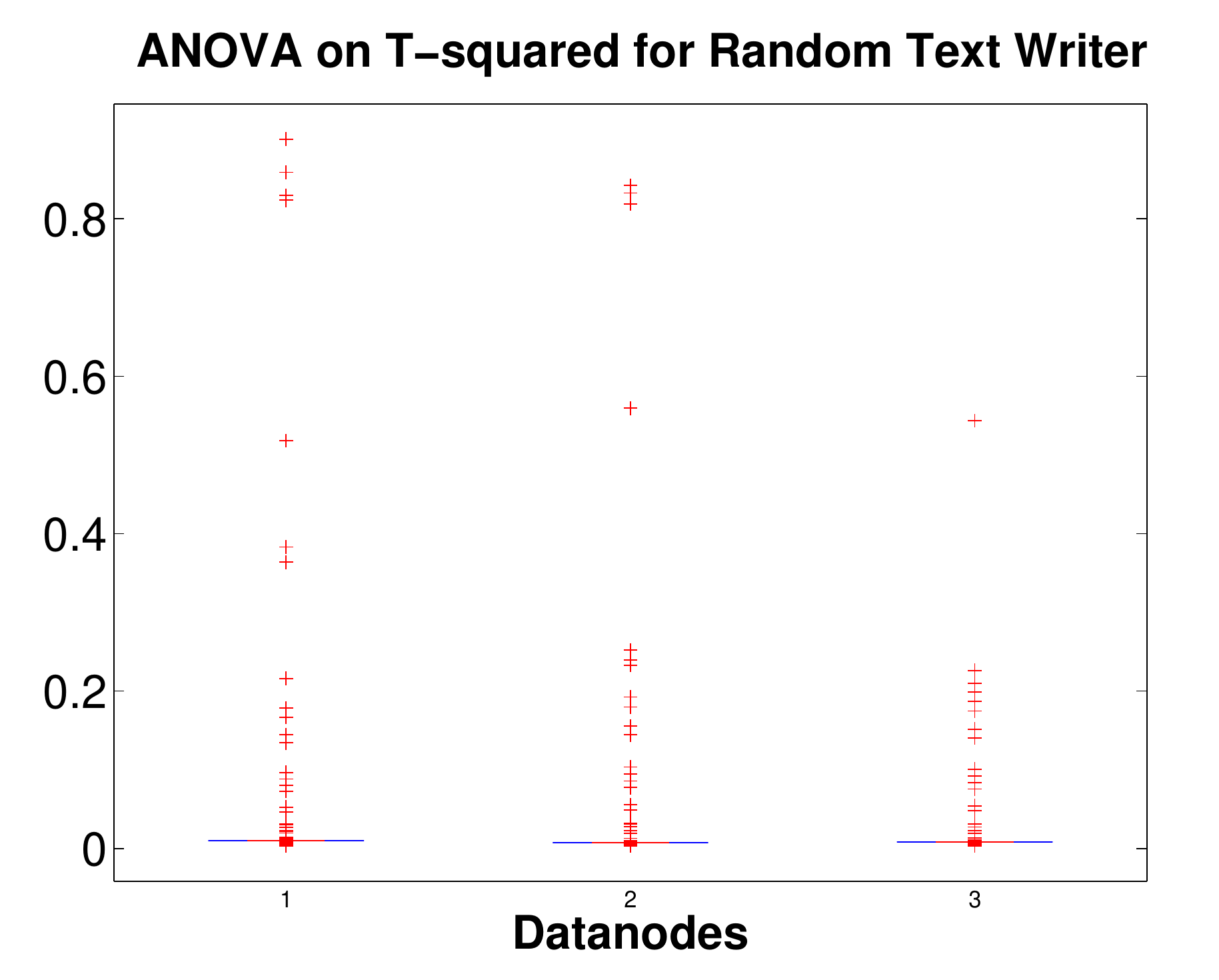}
        \caption{Random Text Writer}
        \label{fig:anova-rtw}
    \end{subfigure}
        ~ 
    \begin{subfigure}[b]{0.23\textwidth}
        \includegraphics[trim = 15mm 70mm 15mm 60mm, clip, width=\textwidth, height=5cm]{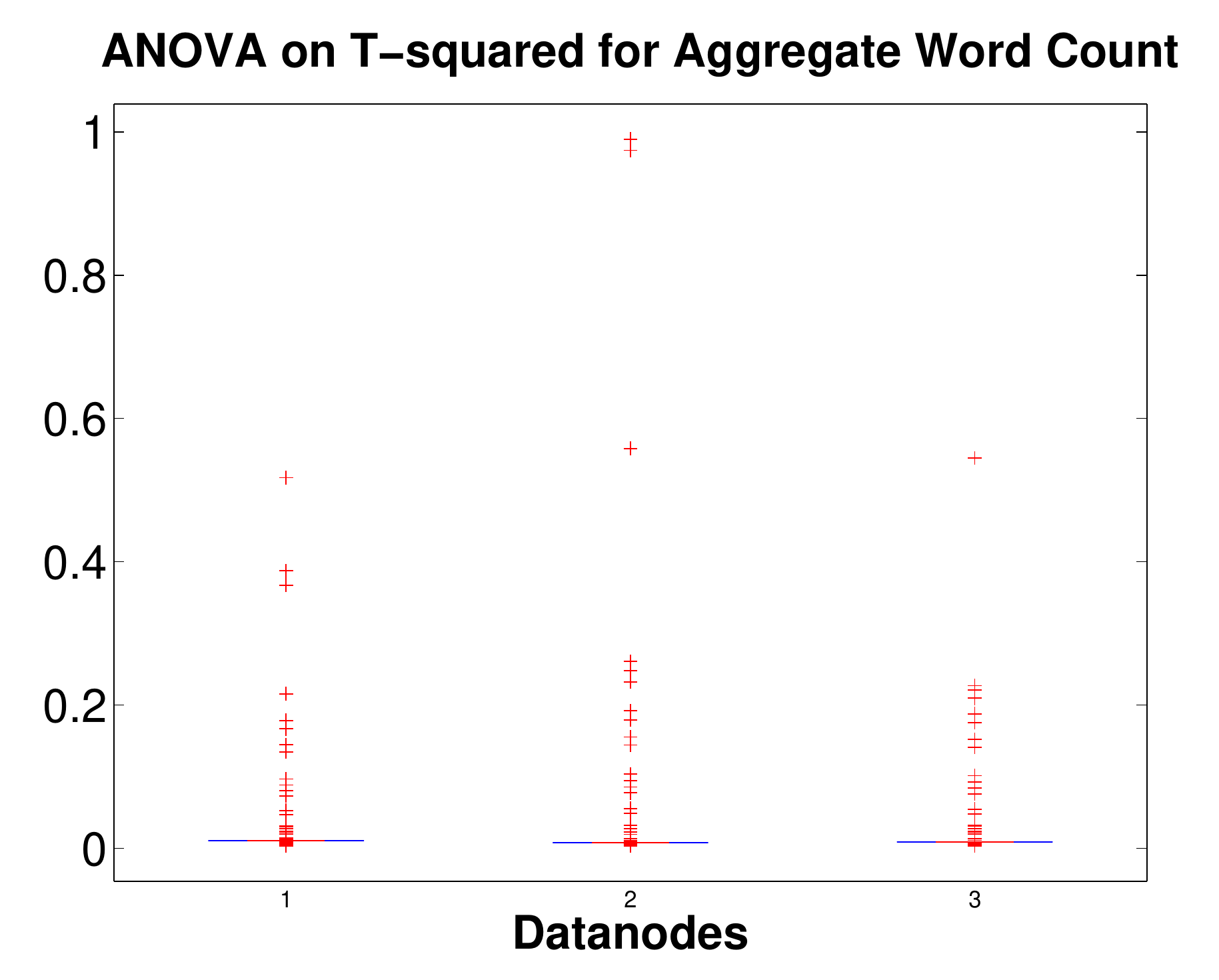}
        \caption{Aggregate Word Count}
        \label{fig:anova-awc}
    \end{subfigure}
    \caption{Results showing difference in Memory Mappings ANOVA on $t^2$ of 3 datanodes after a replica was compromised.}
    \label{fig:attack1-mem-anova}
\end{figure*}

\begin{figure*}[!t]
    \begin{subfigure}[b]{0.23\textwidth}
        \includegraphics[trim = 15mm 60mm 15mm 60mm, clip, width=\textwidth, height=5cm]{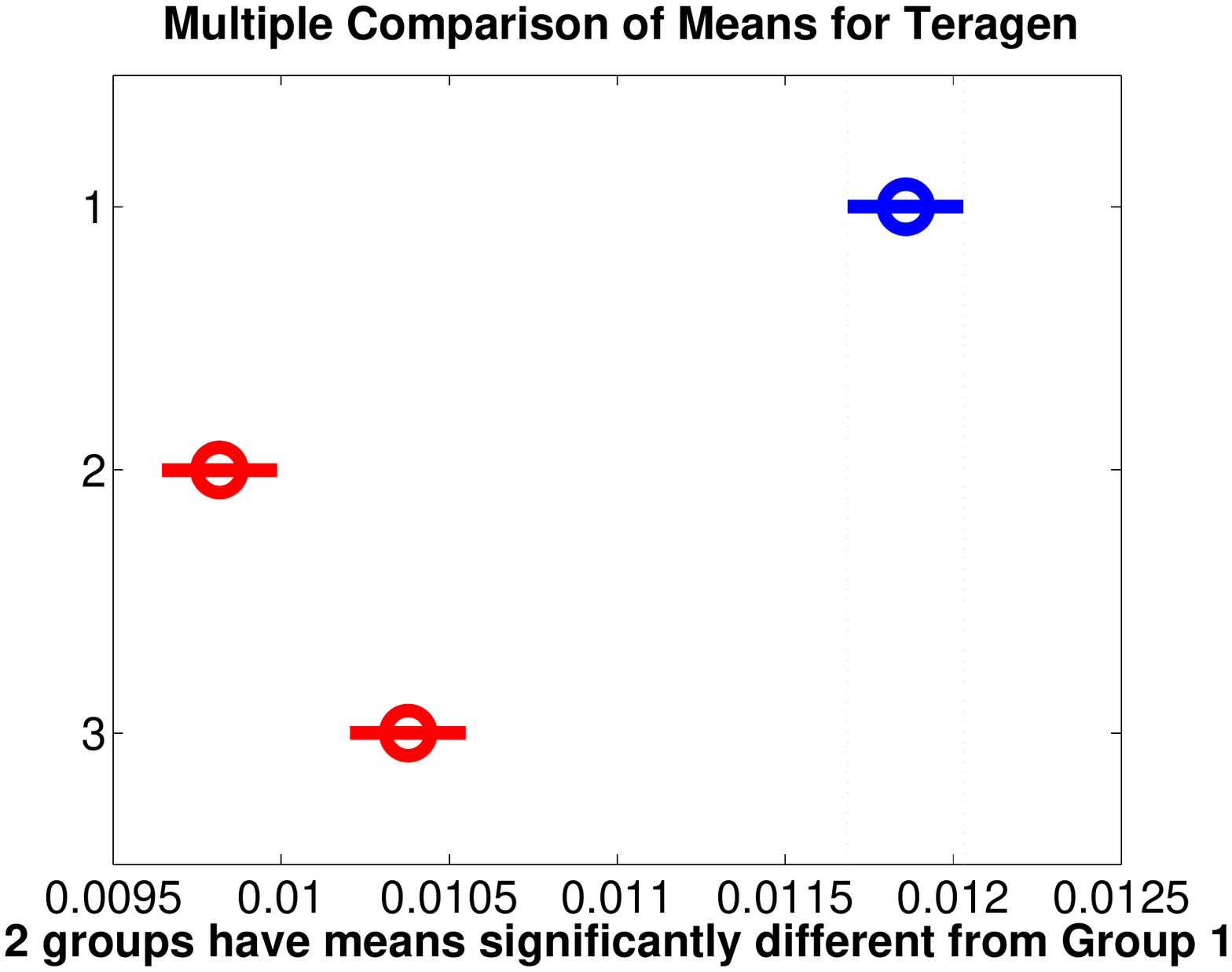}
        \caption{Teragen}
        \label{fig:hsd-tg}
    \end{subfigure}
    ~ 
    \begin{subfigure}[b]{0.23\textwidth}
        \includegraphics[trim = 15mm 60mm 15mm 60mm, clip, width=\textwidth, height=5cm]{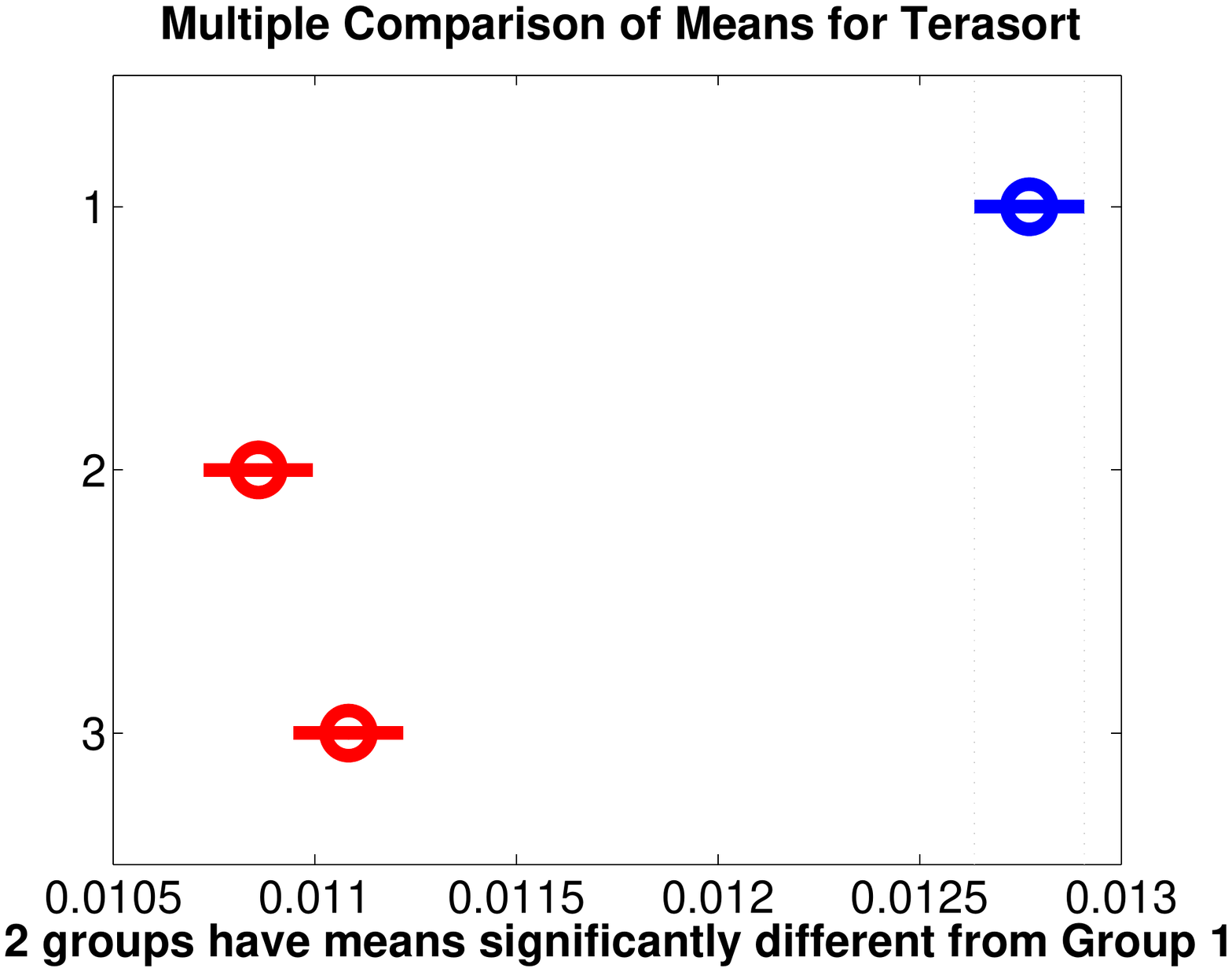}
        \caption{Terasort}
        \label{fig:hsd-ts}
    \end{subfigure}
    ~ 
    \begin{subfigure}[b]{0.23\textwidth}
        \includegraphics[trim = 15mm 60mm 15mm 60mm, clip, width=\textwidth, height=5cm]{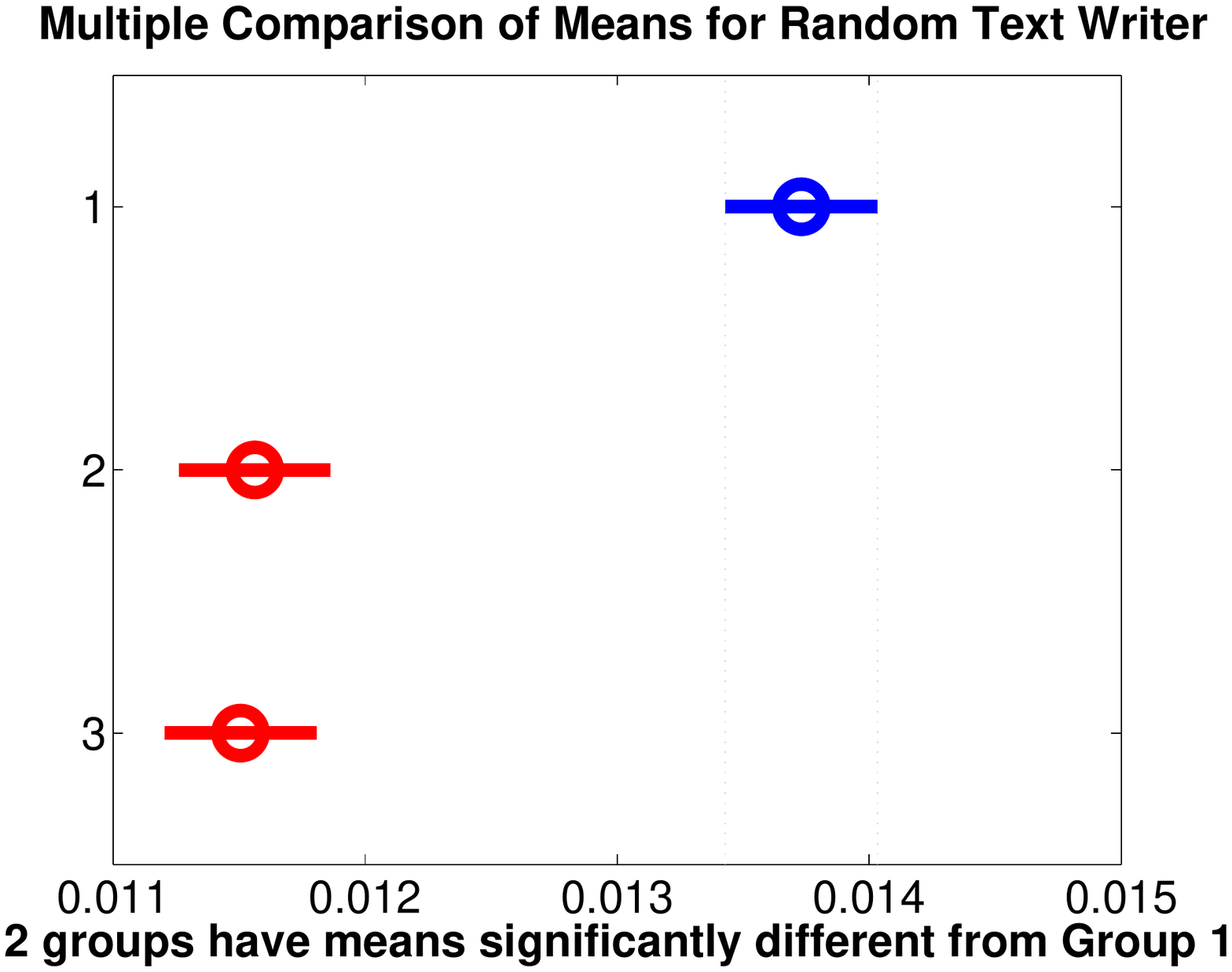}
        \caption{Random Text Writer}
        \label{fig:hsd-rtw}
    \end{subfigure}
        ~ 
    \begin{subfigure}[b]{0.23\textwidth}
        \includegraphics[trim = 15mm 60mm 15mm 60mm, clip, width=\textwidth, height=5cm]{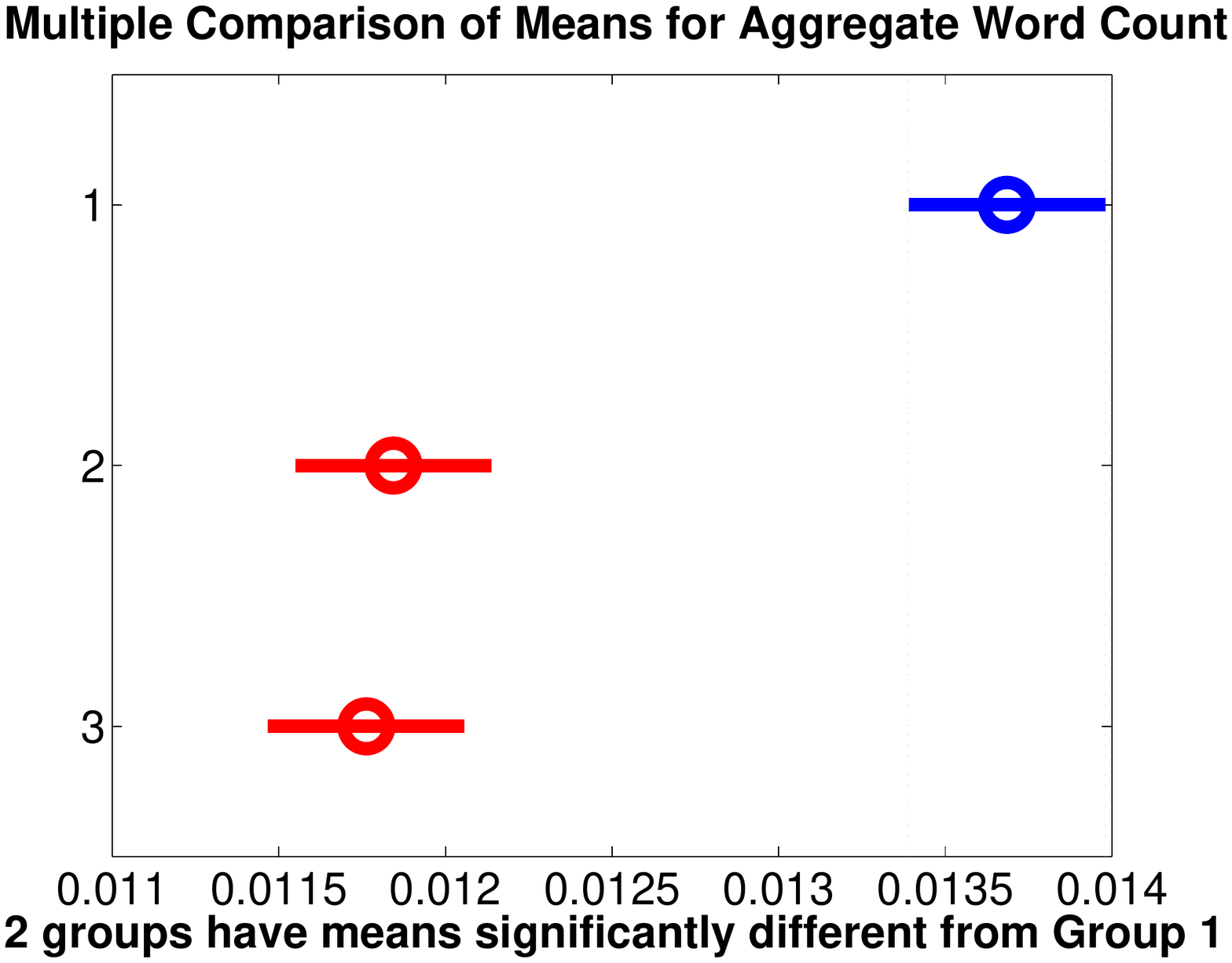}
        \caption{Aggregate Word Count}
        \label{fig:hsd-awc}
    \end{subfigure}
    \caption{Results showing difference in Memory Mappings Tukey on $t^2$ of 3 datanodes after a replica was compromised.}
    \label{fig:attack1-mem-tukey}
\end{figure*}

\begin{figure*}[!t]
    \centering
    \begin{subfigure}[b]{0.23\textwidth}
        \includegraphics[trim = 5mm 60mm 5mm 60mm, clip, width=\textwidth, height=4.5cm]{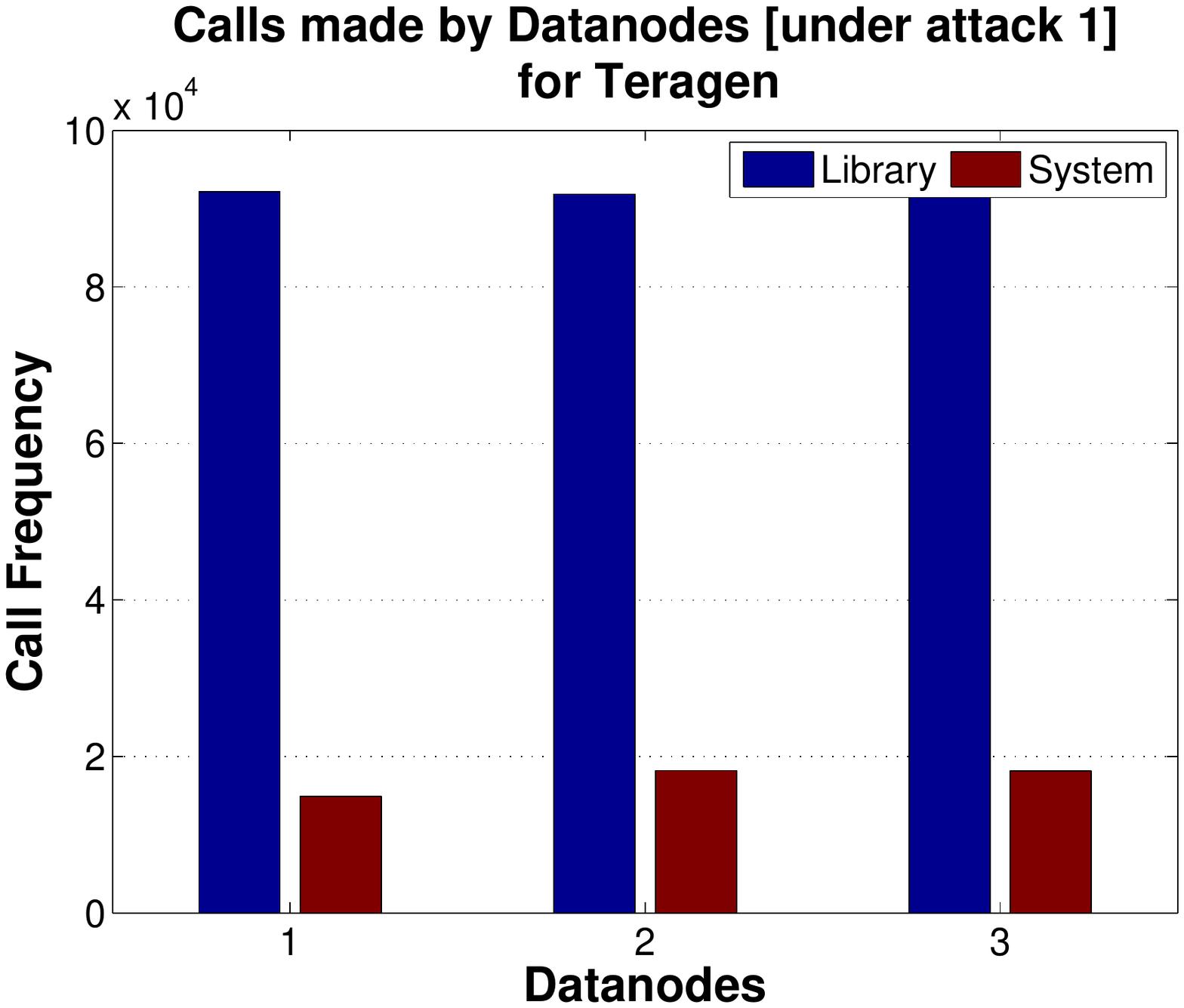}
        \caption{Teragen}
        \label{fig:call-tg}
    \end{subfigure}
    ~ 
    \begin{subfigure}[b]{0.23\textwidth}
        \includegraphics[trim = 5mm 60mm 5mm 60mm, clip, width=\textwidth, height=4.5cm]{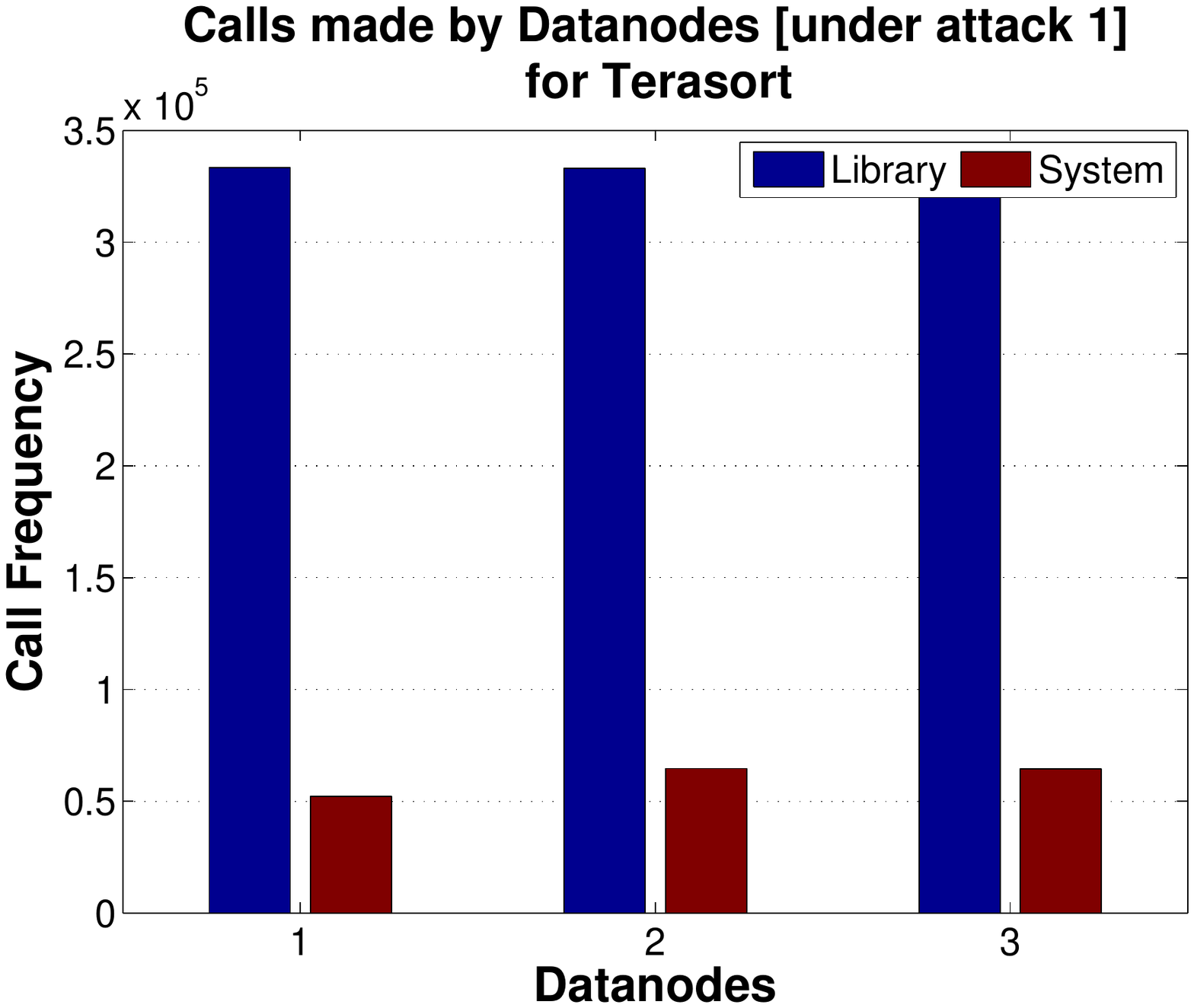}
        \caption{Terasort}
        \label{fig:call-ts}
    \end{subfigure}
    ~ 
    \begin{subfigure}[b]{0.23\textwidth}
        \includegraphics[trim = 5mm 60mm 5mm 60mm, clip, width=\textwidth, height=4.5cm]{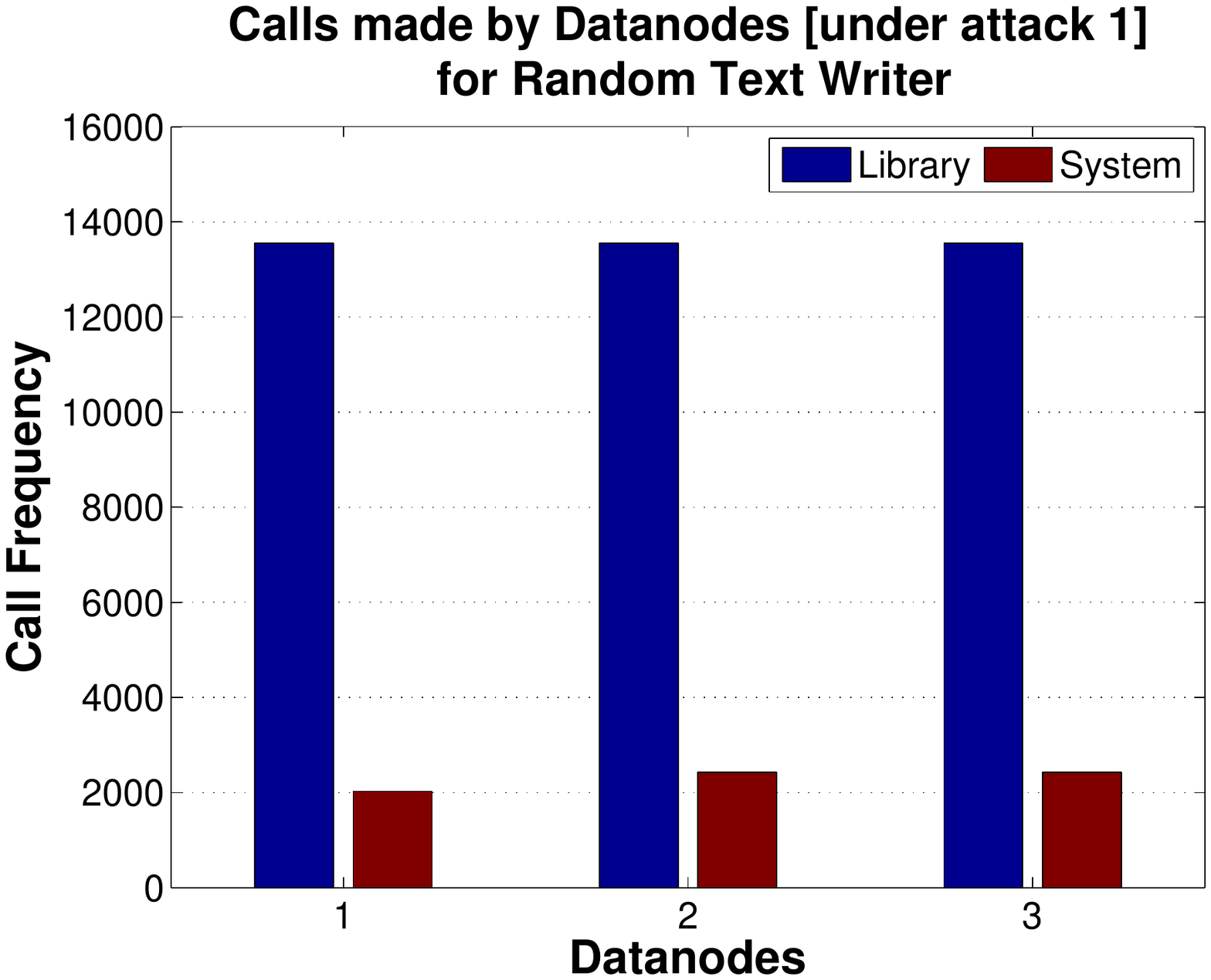}
        \caption{Random Text Writer}
        \label{fig:call-rtw}
    \end{subfigure}
        ~ 
    \begin{subfigure}[b]{0.23\textwidth}
        \includegraphics[trim = 5mm 60mm 5mm 60mm, clip, width=\textwidth, height=4.5cm]{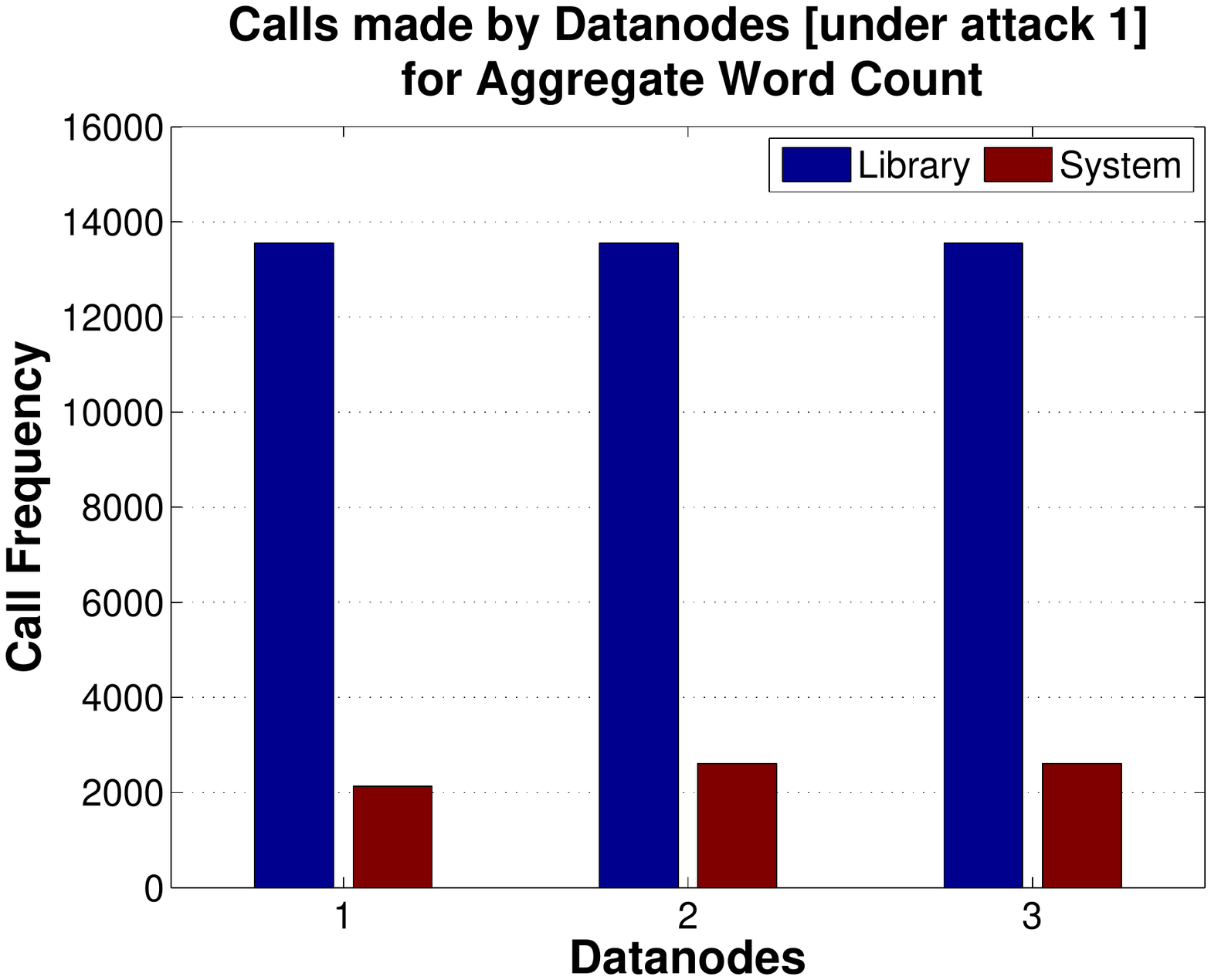}
        \caption{Aggregate Word Count}
        \label{fig:call-awc}
    \end{subfigure}
    \caption{Results showing similarity in Calls of 3 datanodes after a replica was compromised by attack 1.}
    \label{fig:attack1-call}
\end{figure*}

\textit{Attack 1 Results (Modifying a datanode configuration)}: It can be noticed from Figures \ref{fig:teragen} and \ref{fig:terasort} that the Amazon EC2 cluster monitoring metrics such as CPU utilization, Network traffic (bytes in and out) were unable to detect the insider attack while running the Terasort and Teragen examples. But the results from our method for the same Hadoop examples clearly indicate that there is an intrusion in the system, as noticed in Figures \ref{fig:attack1-mem-anova} and \ref{fig:attack1-mem-tukey}. ANOVA on the $t^2$ vectors from the datanodes indicates that one of the datanodes has a different distribution compared to the other two. This can be observed in the $p-value$ column of Table \ref{table_basic_results}. In all four examples, the $p-value$ is extremely low and indicates strong rejection of the null hypothesis that the means of the three different distributions are similar. The multiple comparison test proves that the means of these distributions are not equal and that datanode 1 (in blue) is the one that is different from the other two datanodes (in red). Figures \ref{fig:anova-tg} - \subref{fig:anova-awc} show the results of ANOVA and Figures \ref{fig:hsd-tg} - \subref{fig:hsd-awc} show the results of multiple comparison test. Interestingly, the call frequency on all nodes for these examples seemed to follow similar patterns and the number of distinct library calls made by a datanode is always constant. So if we just consider call frequency analysis for threat detection, this attack is an example of false positive. But it is the system call frequency that hints at the possibility of an attack. Since the memory size and the number of threads for datanode1 were reduced and compared to the other two datanodes, it can be noticed that the system calls (calls to the stack) are relatively low for datanode1 in all examples. This can be observed in Figure \ref{fig:attack1-call}. 

\begin{figure*}[!t]
    \centering
        \begin{subfigure}[b]{0.47\textwidth}
        \includegraphics[trim = 5mm 60mm 5mm 60mm, clip, width=\textwidth]{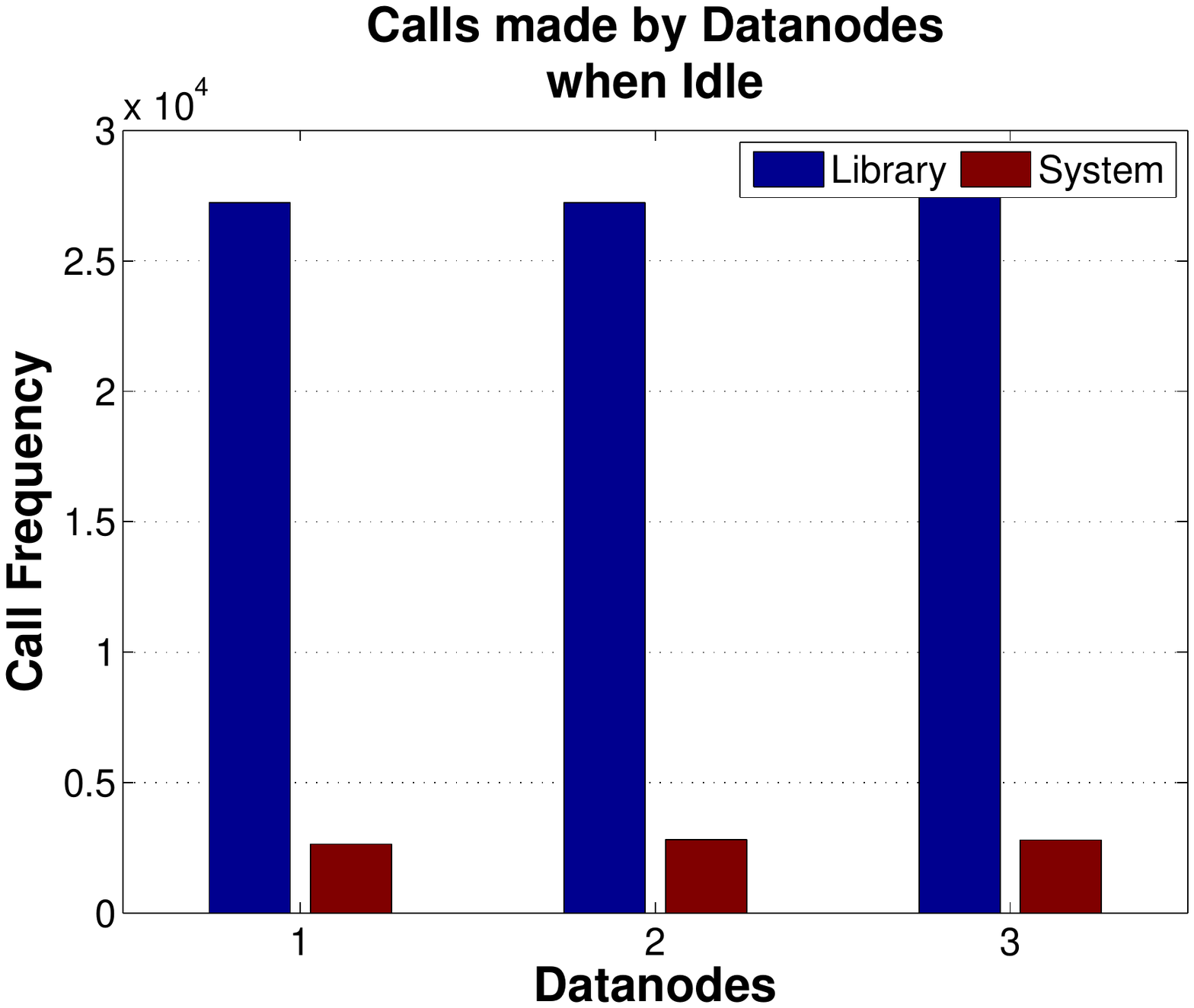}
        \caption{Calls made when \\ cluster is idle}
        \label{fig:call-idle}
    \end{subfigure}
 ~ 
    \begin{subfigure}[b]{0.47\textwidth}
        \includegraphics[trim = 5mm 60mm 5mm 60mm, clip, width=\textwidth]{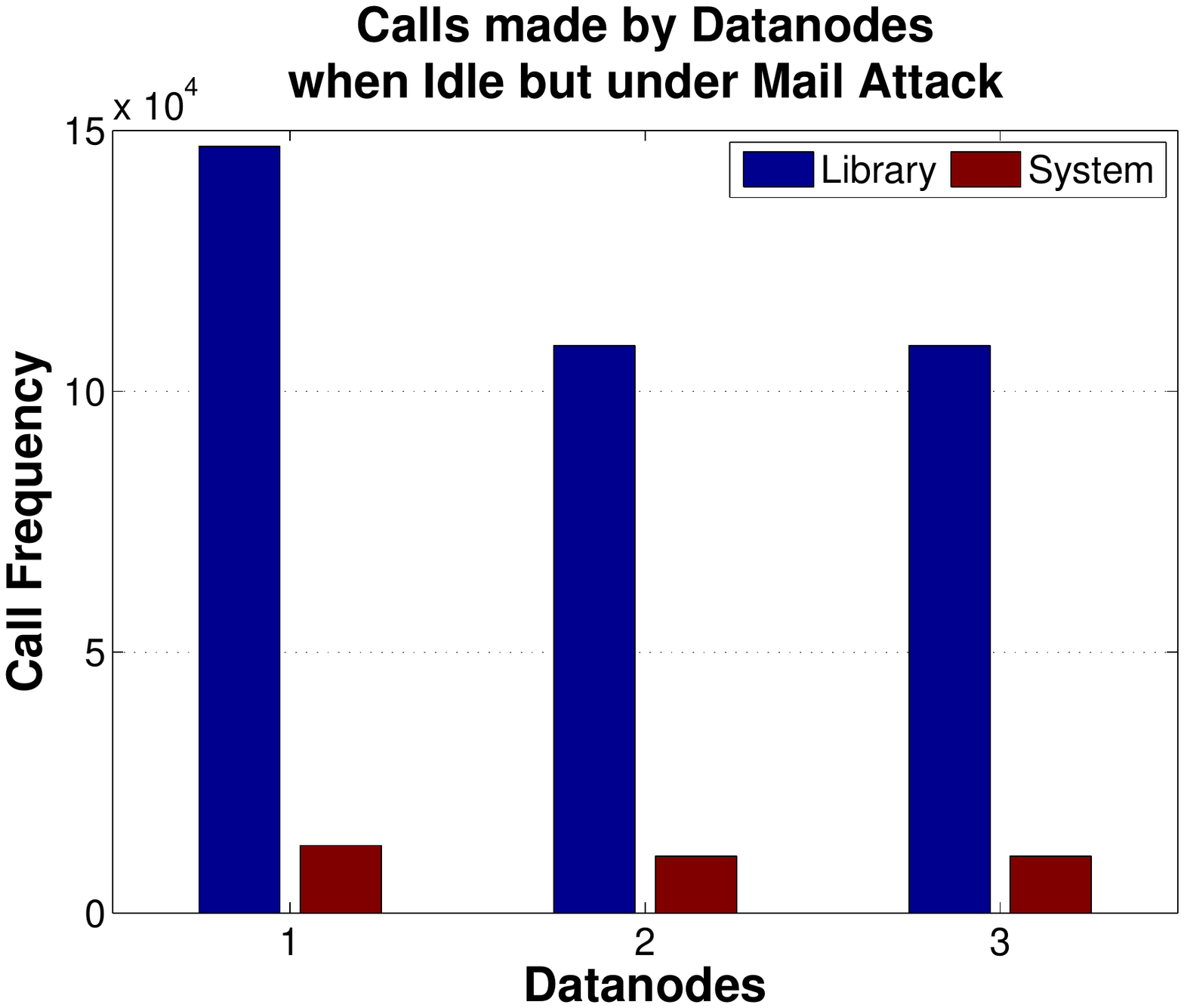}
        \caption{Calls made when cluster is idle but with \textbf{mail attack} in the background}
        \label{fig:call-idle-mail}
    \end{subfigure}
    \caption{Results showing difference in calls from 3 datanodes after a replica was compromised by attack 2.}
    \label{fig:compromised-call}
\end{figure*}

\textit{Attack 2 Results (Illicit copying of data)}: Since our test setup uses Amazon EC2, we cannot use a USB drive to copy files. Instead we tried to access data from the $\texttt{/dev}$ folder because all nodes in the cluster are running on Linux operating system. It must be noted that for this kind of attack, it is not required to perform an action (run an example) to notice that the system has been compromised. Hence, this analysis is performed when the system is idle. A script used for encrypting and sending files in RAM disks as mail attachments to system admin's personal email account. Each file is 4MB in size and it is zipped before sending out as mail attachment. This leads to a difference in the call frequency pattern of the datanode, as observed in Figure \ref{fig:compromised-call}. It can be observed from the call frequency in Figures \ref{fig:call-idle} and \ref{fig:call-idle-mail} that compromised datanode i.e. datanode1's call frequency is order of magnitude more when compared to datenode2 and datanode3 which were not compromised. 

\subsection{Analysis and Observations}
The time complexity of PCA is O($p^{2}n+p^{3}$) where n is the number of observations and p is the number of variables in the original dataset. In our case, $p = 3$ and even if we generalize, the value of $p$ will be some constant \textit{k} because it represents the number of features in memory to be observed. Also, this constant $k$ will be much smaller than $n$. So, the time complexity of PCA in our case should be approximately O($n$) i.e., linearly dependent on the number of observations made. 

In case of memory pattern analysis, if the tails in the observed populations have considerably larger values compared to the mean of the non-tail data, then those data points will have an impact on the output of variance analysis tests such as ANOVA. Hence, it is important to first filter out such data points before running the analysis test. In case of call analysis, there cannot be a concrete conclusion about the system being attacked based only on frequency of calls obtained from different datanodes. Hence, one can conclude that a combination of both of these methods along with other traditional security methods is needed to keep the system safe.

Intrusion detection methods need to account for the following:
\begin{enumerate}[(1)]
\item \textit{True Positive:} successful identification of anomalous or malicious behavior. Our framework achieves this for all data attacks because accessing data involves memory allocation or reference.
\item \textit{True Negative:} successful identification of normal or expected behavior. Our framework achieved this when tested on idle datanodes.
\item \textit{False Positive:} normal or expected behavior is identified as anomalous or malicious. Our framework will have this problem if the the memory mapping observations are not properly cleaned (as mentioned above). A false positive in our framework might also occur when there is a delay in the communication among datanodes about the profile.
\item \textit{False Negative:} anomalous or malicious behavior should have been identified but the framework could not. This case arises if the all duplicate datanodes in the big data cluster are attacked by an insider at once. Luckily, this is highly unlikely to happen in case of large, distributed big data clusters. Other traditional security mechanisms in place will be able to prevent such cases from happening \cite{hadoop,spark}.
\end{enumerate}

\section{Conclusion}
This work is a full implementation of our previous proposal \cite{santosh4}. In this work, we propose a technique to mitigate vulnerabilities and detect attacks during runtime within big data platforms. Our proposed technique analyzes system \& library calls along with memory accesses of a process, packs all of the analysis information together as a process behavior profile and shares that profile with other replica datanodes in the system. The replica datanodes verify the received call traces and access patterns with their local processes for attack detection. Experimental results show that our approach can detect insider attacks even in cases where the usual CPU and network analysis fail to do so, when tested on Hadoop MapReduce examples.    

For future work, we would like to explore security and privacy of user data for big data platforms that specifically support machine learning algorithms and differential privacy. 

%\newpage

\section*{Acknowledgment}
The authors would like to thank Amazon EC2 for their research grant. The authors would also like to thank Ali Reza Chakeri, Fillipe de Souza and Tahoma Toelkes for their valuable inputs.

\section*{References}
\bibliography{mybibfile}

\appendix

\section{Changing the configuration of HDFS on a hadoop datanode} \label{App:AppendixA}
The \texttt{hdfs-site.xml} file on a datanode has the HDFS configuration details. With hadoop distribution version 2.7.1 there are 20 properties that can be set for HDFS on a datanode. These include details about threads, read/write policies, memory \& cache allocations and their policies. Changing the contents of this file on one of the datanodes is a insider job that affects the performance of the overall cluster but cannot be detected easily. For this work, we played around with some of the HDFS properties such as having an additional local folder to save all data processed by the datanode, reducing the size of memory allocated for HDFS on the datanode, changing the thread handler count used by the datanode and drastically decreasing the revocation polling time for the datanode.

\section{Using external devices on Linux} \label{App:AppendixC}
Successfully opening a device file name (e.g. /dev/xvda1, /dev/loop0) establishes a link between a file descriptor and an attached device. In case of Amazon EC2 nodes, we typically find Xen disk storage devices, RAM disks or loop devices because it is a virtual machine. Value of a symbolic link to a device can be obtained using a command such as: $\texttt{sudo readlink -f /dev/input/by-path/*}$ to access input devices, $\texttt{sudo readlink -f /dev/loop*}$ to access loop devices,
$\texttt{sudo readlink -f /dev/ram*}$ to access ram disks. These symbolic links can be used by most system calls. Commands such as $\texttt{dd}$ can be used to convert and copy these files. The associated driver maintains state information and resources at both the device and the file descriptor levels. The possible system calls are: ioctls(); open(); write(); read(); poll(); close(); mmap(); and fcntl(). The number of possible error codes are also limited.

%\section{Calculating Sum-of-squares with 3 latent variables} \label{App:AppendixD}
%Sum of squares calculates the sum of squared differences from the mean. Basically, it calculates the variance. If we have multiple variables, lets say $x,y,z$ and set of observations for each variable then we can calculate the $t^2$ statistic using the formulas below. Here, $t^2$ values represent a measure of the variation in each sample within the model. It indicates how far each sample is from the center of the model. Relative contributions show how two or more samples differ from each other and this can be calculated numerically with simple subtraction. 

\end{document}